\documentclass[aps,pra,twocolumn,amsmath,amssymb,superscriptaddress,showpacs]{revtex4}
\usepackage{graphics}
\usepackage{graphicx}
\usepackage{bbm}
\usepackage{amsfonts}
\usepackage{amssymb}
\usepackage{epsfig}

\begin{document}

\title{Emergent phases in a compass chain with multi-site interactions}

\author {Wen-Long You}
\affiliation{\mbox{College of Physics, Optoelectronics and Energy, Soochow
University, Suzhou, Jiangsu 215006, P.R. China}}

\author {Cheng-Jie Zhang}
\affiliation{\mbox{College of Physics, Optoelectronics and Energy, Soochow
University, Suzhou, Jiangsu 215006, P.R. China}}

\author {Weihai Ni}
\affiliation{\mbox{College of Physics, Optoelectronics and Energy, Soochow
University, Suzhou, Jiangsu 215006, P.R. China}}

\author {Ming Gong}
\affiliation{\mbox{Key Laboratory of Quantum Information and
Synergetic Innovation Center of Quantum Information and Quantum Physics,} \\
University of Science and Technology of China, Hefei, Anhui 230026, P.R. China}

\author {Andrzej M. Ole\'s }
\affiliation{Max Planck Institute for Solid State Research,
             Heisenbergstrasse 1, D-70569 Stuttgart, Germany }
\affiliation{\mbox{Marian Smoluchowski Institute of Physics, Jagiellonian
             University, prof. S. \L{}ojasiewicza 11, PL-30348 Krak\'ow, Poland}}

\begin{abstract}
We study a dimerised spin chain with biaxial magnetic interacting ions
in the presence of an externally induced three-site interactions out
of equilibrium. In the general case, the three-site interactions play
a role in renormalizing the effective uniform magnetic field. We find
that the existence of zero-energy Majorana modes is intricately
related to the sign of Pfaffian of the Bogoliubov-de Gennes Hamiltonian
and the relevant $Z_2$ topological invariant. In contrast, we show that
an exotic spin liquid phase can emerge in the compass limit through a
Berezinskii-Kosterlitz-Thouless (BKT) quantum phase transition. Such
a BKT transition is characterized by a large dynamic exponent $z=4$,
and the spin-liquid phase is robust under a uniform magnetic field.
We find the relative entropy and the quantum discord can signal the
BKT transitions.  We also uncover a few differences in deriving the
correlation functions for the systems with broken reflection symmetry.
\end{abstract}

\date{\today}

\pacs{73.21.-b,71.10.Pm,78.40.Kc}

\maketitle

\section{Introduction}

Several intriguing phenomena in condensed matter systems originate
from the interplay of strong electron correlations and frustration.
Frustration occurs intrinsically in the systems with degenerate and
partly occupied orbitals. A representative model which stands for the
orbital-orbital interactions in Mott insulators is the so called
two-dimensional (2D) compass model \cite{Nus15} where nearest neighbor
interactions like $\propto\sigma^\alpha_i\sigma^\alpha_j$ (with
$\alpha=x,z$ being spin component) compete with each other along two
different spatial directions of the bonds. In such a frustrated quantum
system, the spins cannot order simultaneously to minimize all local
interactions, and the ground state is highly degenerate.

The dominating finite-range interactions in many-body systems can lead
to the onset of self-ordered phases in spin systems. One-dimensional
(1D) quantum models are natural playgrounds for hosting different
orders and distinct universality, especially some exactly solvable
models such as the 1D compass model \cite{Brz07}. Here again the
competing interactions are between nearest neighbors.
However, the range of the hybridization of the electron wave function
will be more extended in reality than only to nearest neighbor sites
in some realistic bonding geometries, such as CsCoCl$_3$ \cite{Goff95},
LiCu$_2$O$_2$ \cite{Masuda04,Rusydi08}, NaCu$_2$O$_2$ \cite{Capogna10}.
The ramifications are that longer-range interactions should be taken
into account. Complex interactions including the three-spin
interactions between three consecutive sites essentially enrich the
ground state phase diagram of the spin model. Recently three-site
interactions received considerable attention in a bit diverse context.
It was realized that the three-site spin interaction can be included to
exhibit the double ferroic order and multiferroics \cite{Suzuki71}.
Experimentally, systems described by spin-1/2 Hamiltonians with
three-spin interactions can be generated using optical lattices or in
NMR quantum simulators \cite{Tseng99,Peng09,Zhang06}.

So far, much attention has been focused on studies of spin-1/2
isotropic XY (or XX) model chains with two types of three-site
interactions. One is the (XZX$+$YZY)-type of three-site interactions \cite{Eloy12,Derzhko11,Men15,Topilko12,Titvinidze03,Kro08},
where the exchange interaction for next nearest neighbor sites takes on
XX form. The other form of three-site interactions is the (XZY$-$YZX)
type \cite{Men15,Topilko12,Kro08,Lou04}. It has been proven that the XX
chain with the (XZX$+$YZY)-type of three-site interactions can be
transformed to the one with the (XZY$-$YZX)-type and
Dzyaloshinskii-Moriya (DM) interaction by a local spin rotation
\cite{Kro08,Derzhko11}.

On the other hand, a few works have been devoted to investigating the
effects of three-site interactions for anisotropic XY chains, which can
turn to Ising limit and XX limit by varying the anisotropy parameter.
The three-site interactions include again either (XZX$+$YZY) \cite{Cheng10b,Li11,Zvyagin09,Lian11,Cheng10c,Lian11b,Zhang15} or
(XZY$-$YZX) forms \cite{Lou05,Cheng10a,Lei15,Lian11b,Liu12}.
Differently from the situation on the XX chains, two kinds of
three-site interactions on the XY chains are not unitary equivalent.
The (XZX$+$YZY)-type interactions violate the time reversal (T)
symmetry but preserve the parity (P) symmetry, while the
(XZY$-$YZX)-type of three-site interactions break both symmetries
simultaneously. Furthermore, a simplified version of three-site XZX
interactions was also examined \cite{Kopp05,Niu12,Rajak07,Dong16}.
One finds that transverse Ising model with XZX-type interactions is
dual to the XY
model through a nonlocal dual transformation \cite{Fradkin78} which
hosts a number of Majorana zero modes of an open chain \cite{Niu12}.

The organization of the paper is as follows. In Sec. \ref{sec:model}
we introduce the Hamiltonian of the 1D generalized compass model (GCM)
with three-site interactions. Notation is introduced in Sec.
\ref{sec:1D} and then we present the procedure to solve it exactly by
employing Jordan-Wigner transformation.
Ground state properties and excited states are derived in
Sec. \ref{sec:qp}. Majorana modes and topological phase transition are
addressed in Secs. \ref{sec:maj} and \ref{sec:z2}. The exact solution
explains the nature of the quantum phase transition (QPT)
as explained in Sec. \ref{sec:qpt}. The model in the magnetic field is
analyzed in Sec. \ref{sec:field}.  In Sec. \ref{sec:qi} we discuss the
aspects related to quantum information and present the fidelity
susceptibility in Sec. \ref{sec:fs} and coherence susceptibility in
Sec. \ref{sec:coh}. A final discussion and conclusions are presented
in Sec. \ref{sec:summa}.
More technical aspects of the presented solution are given in
Appendices \ref{sec:appa} and \ref{sec:appb}.

\section{The Model and ITS SOLUTION}
\label{sec:model}

\subsection{Generalized compass model in one dimension}
\label{sec:1D}

The 1D GCM  is a microscopic model to mimic
~zigzag spin chains in perovskite transition metal (TM) oxides. For
instance, Co$^{2+}$ ions in CoNb$_2$O$_6$ compound form zigzag chains
along the $c$ axis. At low temperatures, Co spins orient
themselves along two different easy axes in the nearly $(a,c)$ plane
with a 31$^\circ$ canting angle from the $c$ axis. The Peierls-type
spin-phonon coupling renders frustrated spin exchanges along distorted
TM-O-TM bonds \cite{Mochizuki11}.

\begin{figure}[t!]
 \includegraphics[width=8cm]{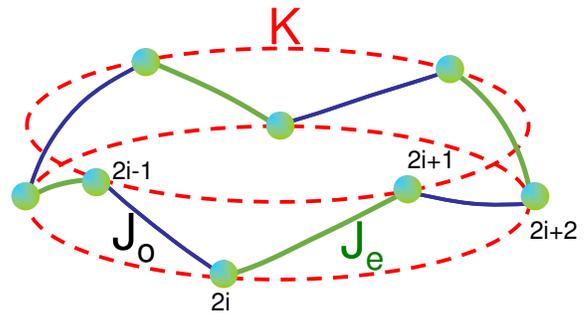}
\caption{ Schematic representation of a zigzag spin chain with periodic
boundary conditions described by the Hamiltonian Eq. (\ref{model}).
Nearest neighbor exchange interactions alternate between $J_o$ (blue thin
lines) and $J_e$ (green thick lines) on odd and even bonds, respectively.
$K$ denotes three-site exchange parameter  (red dashed lines).
}
\label{Fig:Schematic}
\end{figure}

The 1D GCM with alternating exchange interaction considered below
is given by \cite{You1,You2,You16,Wu17},
\begin{eqnarray}
\cal{H}_{\rm GCM}&=& \sum_{i=1}^{N^\prime}
J_{o}\tilde{\sigma}_{2i-1}(\theta)\tilde{\sigma}_{2i}(\theta)
+J_{e}\tilde{\sigma}_{2i}(-\theta)\tilde{\sigma}_{2i+1}(-\theta),
 \nonumber \\
 &+& \sum_{i=1}^{N'}\left( \vec{h}_{o}\cdot\vec{\sigma}_{2i-1}
 + \vec{h}_{e}\cdot\vec{\sigma}_{2i}\right),
\label{Hamiltonian1}
\end{eqnarray}
where the operator with a tilde sign is defined as a linear
combinations of $\{\sigma_i^x,\sigma_i^y\}$ pseudospin components,
\begin{eqnarray}
\tilde{\sigma}_{i}(\theta)&\equiv& \cos(\theta/2)\,\sigma_{i}^x
+\sin(\theta/2)\,\sigma_{i}^y.
\end{eqnarray}
Here, $N'=N/2$ is the number of two-site unit cells. $J_o$ ($J_e$)
denotes the amplitude of the nearest-neighbor planar interaction on odd
(even) bonds, while $\vec{h}_{o}$ ($\vec{h}_{e}$) is the magnitude of
the external field exerted on odd (even) sites. In addition, effective
(XZX$+$YZY)-type three-site interactions are also taken into account,
\begin{eqnarray}
{\cal H}_{\rm 3} &=&K\sum_{i=1}^{N}
\left(\sigma^x_{i-1}\sigma^z_{i}\sigma^x_{i+1}
+\sigma^y_{i-1}\sigma^z_{i}\sigma^y_{i+1}\right),
\label{Hamiltonian2}
\end{eqnarray}
where $K$ characterizes the strength of uniform exchange interaction
between three consecutive spins. Multi-site interactions emerge
simultaneously with two-body interactions as higher-order corrections
in Mott insulating phases, but they are generally believed to have a
negligible effect \cite{Hirsch79}. However, the experimental capability,
such as cold atom technology, allows us to control three-spin
interactions across a wide parameter range \cite{Pachos04}.
Remarkably, three-site interactions appear naturally as an energy
current when a compass chain was in the nonequilibrium steady states
\cite{Qiu16,Robin16}, which can be formally calculated by taking a time
derivative of the energy density and follows from the continuity
equation \cite{Zotos97,Antal97}.
Then the complete Hamiltonian of the 1D GCM with the three-site (XZX$+$YZY)
interactions is,
\begin{eqnarray}
{\cal H}&=& \cal{H}_{\rm GCM} + {\cal H}_{\rm 3}.
\label{model}
\end{eqnarray}
Exchange couplings are shown schematically in Fig. \ref{Fig:Schematic}.
We shall mention that the combined model may be realized by quantum
engineered artificial systems.
For instance, coupling superconducting qubits to microwave circuitry
provide a laboratory to simulate various spin models \cite{Salathe15}
and even multi-site interactions \cite{Mezzacapo14}.
In particular, the cavity array can be driven and dissipative and thus
be settled in a non-equilibrium steady state \cite{Dong16,Bardyn12}.
As the simulated spin chain is driven out of equilibrium in the
presence of an energy current, critical phase transitions between the
pristine ground state and the current-carrying phase and the associated
universality can be probed.

\subsection{Quasiparticles at finite three-site interactions
            and vanishing magnetic field}
\label{sec:qp}

The Jordan-Wigner transformation maps explicitly the pseudospin
operators to spinless fermion operators \cite{lieb61,Katsura62,EBarouch70}:
\begin{eqnarray}
\sigma_j^z&=&1-2c_j^\dagger c_j,               \nonumber \\
\sigma_j^x&=&e^{i\phi_j} (c_j^\dagger + c_j),  \nonumber \\
\sigma_j^y&=&i e^{i\phi_j} (c_j^\dagger - c_j),
\end{eqnarray}
with $\phi_j$ being the phase string generated by all earlier sites
along the chain, $\phi_j=\pi\sum_{l<j}c_l^\dagger c_l^{}$. Neglecting
boundary
terms we arrive at a simple bilinear form of the Hamiltonian
expressed by spinless fermions:
\begin{eqnarray}
{\cal H}&=&
\sum_{j} \left[J_{o}\left(e^{i\theta} c_{2j-1}^{\dagger}c_{2j}^{\dagger}
  +    c_{2j-1}^{\dagger} c_{2j}^{}\right) + J_{e}
  \left(e^{-i\theta} c_{2j}^{\dagger} c_{2j+1}^{\dagger}\right.\right.
\nonumber \\ &+&\!\left.\left.  c_{2j}^{\dagger} c_{2j+1}^{}\right)
+ 2K\left(c_{2j-1}^{\dagger}c_{2j+1}^{}+c_{2j}^{\dagger}c_{2i+2}^{}\right)
+{\rm H.c.}\right].
\label{Hamiltonian2}
\end{eqnarray}
The fermionized version of the model (\ref{Hamiltonian2}) corresponds
to a $p$-wave superconductor in which the electrons have next nearest
neighbor hopping.  There is a relative phase $e^{i\theta}$ between the
nearest neighbor hopping and the nearest neighbor pairing. The present
model is also dual to an extended Su-Schrieffer-Heeger (SSH) model
\cite{Tong15,Jafari17}.

\subsection{Majorana zero modes of topological nontrivial states}
\label{sec:maj}

In this Section we explore the zero modes via the Bogoliubov-de Gennes
(BdG) equations with open boundary condition (OBC).
Generally, Hamiltonian (\ref{Hamiltonian2}) is not PT symmetric except
for $\theta=\pi/2$ when the $p$-wave pairing amplitude is purely
imaginary. It can be diagonalized with a linear transformation of the
canonical fermion operators $\{c,c^\dagger\}$,
\begin{eqnarray}
{\cal H}&=& \sum_{i,j}\left[c_{i}^{\dagger} A_{i j}c_{j}
+\frac{1}{2}\left(c_{i}^{\dagger}B_{ij}c_{j}^{\dagger}
-c_{i}B_{i j}^*c_{j}\right)\right],
\label{ABHam}
\end{eqnarray}
with
\begin{eqnarray}
A_{i j}&=&J_i (\delta_{i,j-1} + \delta_{i,j+1})
 +2K (\delta_{i,j-2} + \delta_{i,j+2}),                      \nonumber \\
B_{i j}&=&\Delta_i(\delta_{i,j-1} - \delta_{i,j+1}),
\label{ABMatrix}
\end{eqnarray}
where $J_i=J_o$ ($J_e$) and $\Delta_i\equiv J_o e^{i\theta}$
($\Delta_i\equiv J_e e^{-i\theta}$) when $i\in$ odd (even).
$A$ ($B$) is a $N\times N$ symmetric (antisymmetric) matrix.
Hamiltonian (\ref{Hamiltonian2}) can be diagonalized by using the
BdG transformation:
\begin{eqnarray}
\eta_n^\dagger=\sum_{i=1}^N\left(u_{n,i} c_i^\dagger +v_{n,i}^* c_i^{}\right),
\end{eqnarray}
where $n$ and $i$ are
indices of eigenvalues and lattice sites,
respectively. The spectra $\Lambda_n$ and eigenvectors $u_{n,i}$ and
$v_{n,i}$ can be determined by solving BdG equations \cite{Derzhko98}:
\begin{eqnarray}
\left(
  \begin{array}{cc}
    A & B \\
    -B^* & -A^T
     \\
  \end{array}
\right)
\left(\begin{array}{c}
    u_{n,i}\\
    v_{n,i}^*
     \\
       \end{array}
     \right)= \Lambda_n \left(\begin{array}{c}
    u_{n,i}\\
    v_{n,i}^*
     \\
       \end{array}
     \right).
\end{eqnarray}

The BdG Hamiltonian satisfies an imposed symmetry, i.e., particle-hole
symmetry (PHS), in the form $\tau^x {\cal H}^T\tau^x =-{\cal H}$,
where the Pauli matrix $\tau^x$ acts in Nambu space. Hence the energy
levels must come in $\pm \Lambda$ conjugate pairs except the zero
energy mode which is self-conjugate.
The topological point defects in the 1D model trap zero-energy bound
states and induce at most one protected zero-energy mode localized at
each end of an open chain. The existence of a zero-energy localized
states can be interpreted as a signature of Majorana modes.

\begin{figure}[t!]
 \includegraphics[width=8cm]{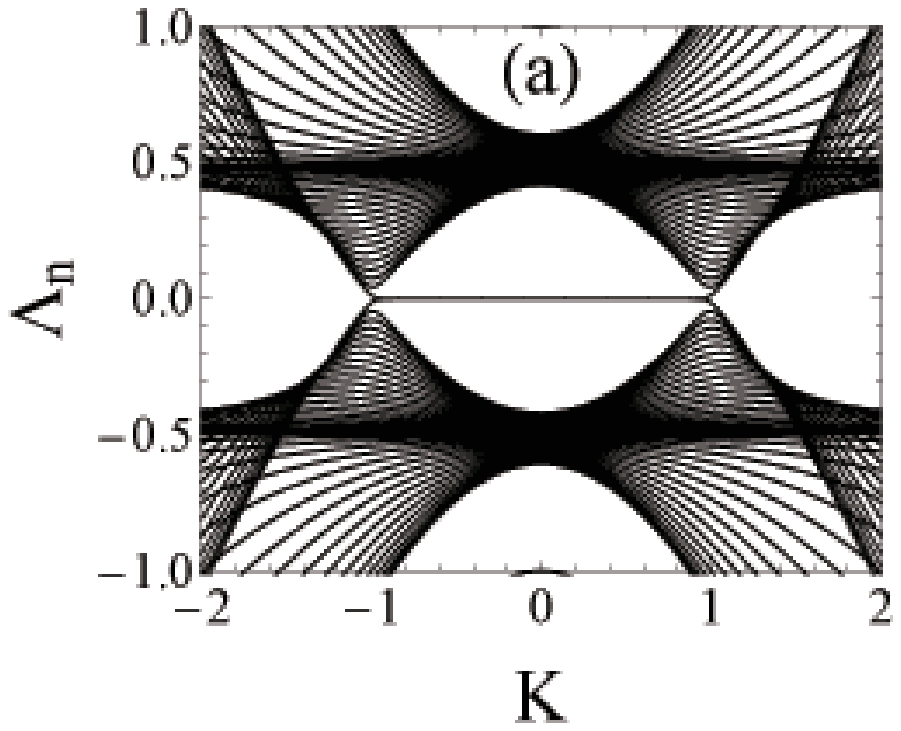}
 \includegraphics[width=8cm]{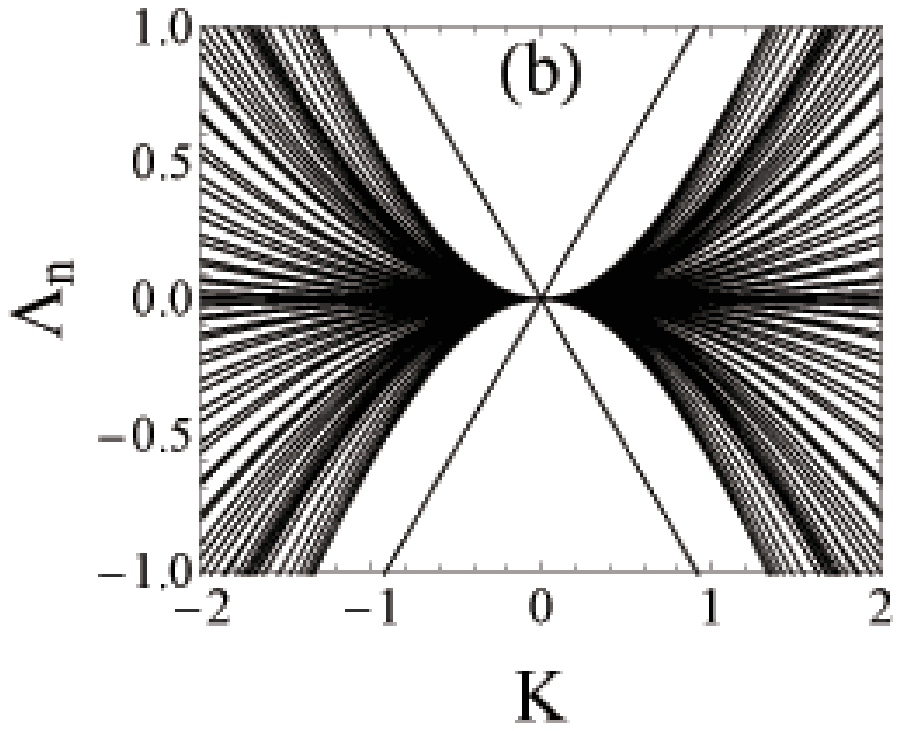}
 \caption{
Energy spectrum for the GCM with the OBC for $N=80$ sites.
The orbital angle $\theta$ is:
(a) $\theta=\pi/3$;
(b) $\theta=\pi/2$.
Parameters are as follows: $J_o$=1, $J_e$=4.}
\label{Fig:Spe_OBC_h=0}
\end{figure}

Figure \ref{Fig:Spe_OBC_h=0} shows the energy spectrum for the GCM
with OBC for two characteristic angles $\theta$. At $\theta=\pi/3$
which is isomorphic with $e_g$ orbital model \cite{Cin10} there
exist zero modes for $\vert K \vert \le 1$ shown in Fig.
\ref{Fig:Spe_OBC_h=0}(a). Extensive data reveals that such edge modes
are protected by energy gaps away from critical points. The model for
$\theta$ $\neq$ $\pi/2$ can be modified continuously to a Kitaev model
in the topological nontrivial phase without closing the band gap, so the
model in such a phase was featured by the presence of zero-energy
Majorana edge states under the OBC, namely, $n_M=1$.
According to Eq. (\ref{Hamiltonian2}), topological phase transitions of
the model are classified in terms of the number of isolated Majorana
zero modes $n_M$. These Majorana states are stable against quadratic
perturbations which preserve the symmetries.

In contrast, there is no zero mode at $\theta$ =$\pi/2$ irrespective of
$K$, see Fig. \ref{Fig:Spe_OBC_h=0}(b).
We note that at $K=0$ there is a macroscopic number $2^{N/2-1}$ of
states condensed at zero-energy modes \cite{Brz07,You08} but they are
not edge modes.
The three-site interactions remove the macroscopic degeneracy instantly
and zero-energy ~states become dispersive. Interestingly, the tower of
these low-energy excitations keep intact as $K$ increases, and they are
separated from higher energy states by a linear dispersion.

\subsection{Pfaffian $\mathbb{Z}_2$ invariant for BdG Hamiltonian
            and topological phase transition}
\label{sec:z2}

Next discrete Fourier transformation for plural spin sites is
introduced for the periodic boundary conditions,
\begin{eqnarray}
c_{2j-1}\!=\frac{1}{\sqrt{N'}}\sum_{k}e^{-ik j}a_{k},\quad
c_{2j}\!=\frac{1}{\sqrt{N'}}\sum_{k}e^{-ik j}b_{k},
\end{eqnarray}
with the discrete momenta as
\begin{eqnarray}
k=\frac{n\pi}{ N^\prime  }, \quad n\!= -(N^\prime-1), -(N^\prime-3),
\ldots,(N^\prime -1).
\end{eqnarray}
Then we write the Hamiltonian in the BdG form in terms of Nambu
spinors,
\begin{eqnarray}
\cal{H} &=&  \sum_{k}
\Gamma_k^{\dagger}
\hat{M}_k
\Gamma_k, \label{FT2}
\end{eqnarray}
where
\begin{eqnarray}
\hat{M}_k%
&=&\frac{1}{2} \left(\begin{array}{cccc}
 F_k &  S_k &  0   &  T_k  \\
S_k^* &  F_k & -T_{-k}    & 0     \\
0 &  -T_{-k}^*    & -F_k & -S_{-k}^*\\
 T_k^*   & 0   &  -S_{-k} &-F_k
\end{array}\right),\label{HamiltonianMatrix2}
\end{eqnarray}
and $\Gamma_k^{\dagger}=(a_{k}^{\dagger},b_k^{\dagger},a_{-k},b_{-k})$ .
The matrix elements in Eq. (\ref{HamiltonianMatrix2}) are:
\begin{eqnarray}
T_k &=& J_o e^{i\theta} - J_e e^{i(k-\theta)}, \label{Tk}\nonumber \\
S_k &=& J_o+J_e e^{ik}, \label{Sk}\nonumber \\
F_k &=& 2 K \cos k. \quad \label{Fk}
\end{eqnarray}
The system belongs to topological class $D$ with a topological invariant
$\mathbb{Z}_2$ in one dimension \cite{Chiu16}, which satisfies
\begin{equation}
{\cal C}^{-1} \hat{M}(-k) {\cal C}=- \hat{M}(k).
\end{equation}
Here
${\cal C}=\tau^x\otimes\sigma^0\cal{K}$, where $\tau^x$ and $\sigma^0$
are the Pauli matrices acting on particle-hole space and spin space,
respectively, and $\cal{K}$ is the complex conjugate operator.

Following the basic definition of particle-hole ${\cal C}$,
an auxiliary function $W(k)= \hat{M}_{4\times 4}(k)\,{\cal C}$ is
defined, and we have $W(k)^T = -W(-k)$. For particle-hole symmetric
momenta $k\in\{0,\pm\pi\}$ in Brillouin zone, we have $W(0)^T = -W(0)$
and $W(\pi)^T = -W(\pi)$, which are both skew matrices. The topology
of the GCM can be characterized by the Pfaffian of the Hamiltonian at
$k=0$ and $\pi$, with $\nu=sgn\{Pf(W(0))Pf(W(\pi))\}$. Here $\nu$ is a
topological protected number, which means that $\nu$ will never change
sign upon deformation as long as the energy gap at $k=0$ and $\pi$ is
not closed. Then $\nu=-1$ (+1) corresponds to the topological nontrivial
(trivial) phase, respectively \cite{Kitaev01,Ghosh10}. The Pfaffian
reads
\begin{eqnarray}
Pf[W(0)]&=&\hskip .3cm 4J_o J_e\cos^2 \theta-4 K^2,   \nonumber \\
Pf[W(\pi)]&=&-4J_o J_e\cos^2 \theta-4 K^2.
\end{eqnarray}
It is easy to find that in the regions
$\vert K\vert\le\sqrt{J_o J_e}\cos \theta$, a topological nontrivial
phase with $\nu=(-1)^{n_M}$ is accompanied with a zero-energy Majorana
mode in Fig. \ref{Fig:Spe_OBC_h=0}.

\section{Quantum Phase Transition}
\label{sec:qpt}

\begin{figure}
\includegraphics[width=8.5cm]{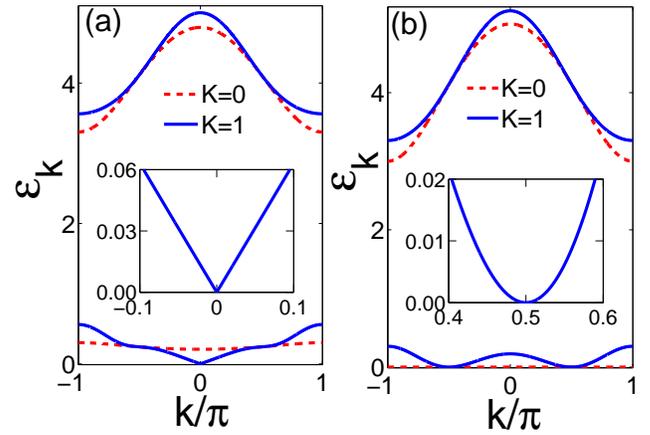}
\caption{Dispersion relations (\ref{excitationspectrum2}) for the GCM
for three selected values of three-spin interactions $K=0$ and
1 at:
(a)~$\theta=\pi/3$ and
(b) $\theta=\pi/2$.
Insets show amplifications of lower energies $\varepsilon_{k,1}$
near Fermi surface at $K=1$. Parameters are as follows: $J_o=1$, $J_e=4$. }
\label{Fig:PBCspectrum}
\end{figure}

Along these lines, we obtain the diagonal form of the Hamiltonian Eq.
(\ref{HamiltonianMatrix2}),
\begin{eqnarray}
{\cal H}=\sum_{k}\sum_{j=1}^{2}  \varepsilon_{k,j}
\left(\gamma_{k,j}^{\dagger}\gamma_{k,j}^{}-\frac{1}{2}\right).
\label{diagonalform}
\end{eqnarray}
The spectra consist of two branches of energies $\varepsilon_{k,j}>0$
($j$= 1,2), given by the following expressions:
\begin{eqnarray}
\varepsilon_{k,1(2)}=\sqrt{\varsigma_k \pm \sqrt{\tau_k }},
\label{excitationspectrum2}
\end{eqnarray}
where
\begin{eqnarray}
\varsigma_k&=&%
\frac{1}{2}(\vert T_k \vert^2 + \vert T_{-k} \vert^2)
+ \vert S_k \vert^2+   F_k^2,   \\
\tau_k&=&
\frac{1}{4}(\vert T_k\vert^2 -\vert T_{-k}\vert^2)^2
+ \vert S_k^* T_k + S_k T_{-k}\vert^2   \nonumber \\
&+& 4 \vert S_k \vert^2 F_k^2.
\end{eqnarray}
Note that $\tau_k$ is always positive for any momentum $k$. We remark
that the energy spectrum $\varepsilon_{k,j}$ $(j=1,2)$ is thus always
positive which is different from the compass spin chain with the
(XZY$-$YZX)-type of three-site interaction \cite{You16}. The form of
$F_k$ (\ref{Fk}) leads to a crossing of excitations at $k=\pm\pi/2$ for
diverse values of $K$, see Fig. \ref{Fig:PBCspectrum}. The most
important properties of the 1D spin system are manifested in the ground
state. The ground-state energy density of our model can be written as
\begin{eqnarray}
e_0 &=& -\frac{1}{2N} \sum_{k} (\varepsilon_{k,1} +\varepsilon_{k,2})
\nonumber \\
&=& -\frac{1}{\sqrt{2N}}\sum_k
\sqrt{\varsigma_k + \sqrt{\varsigma_k^2-\tau_k }}.
\label{E0expression}
\end{eqnarray}

From Eq. (\ref{Hamiltonian2}), $K$ promotes the hopping between next
nearest neighbor sites and modifies the corresponding dispersions.
The phase diagram of the GCM under three-site interactions can be
analytically calculated by investigating the gap closing of the
spectrum (\ref{excitationspectrum2}).
Accordingly, the spectral gap is determined by
the first energy branch, i.e., $\Delta=2 \min_{k}\{\varepsilon_{k,1}\}$.
The gap closes at some critical
momentum $k_c$ delimited by $\varsigma_{k_c}^2$ = $\tau_{k_c}$.
One finds that this condition can be satisfied only when
\begin{eqnarray}
\cos k_c=1 \quad \textrm{and } \quad  K_c=\pm\sqrt{J_o J_e} \cos\theta.
\label{condition3}
\end{eqnarray}
When the magnitude $\vert K\vert$ of three-site interactions
(\ref{Hamiltonian2}) is below the critical field,
$K_c=\sqrt{J_o J_e}\cos\theta$ [see Fig. \ref{Fig:PBCspectrum}(a)], the
ground state is a canted antiferromagnetic phase dominated by nearest
neighbor correlation functions ~along the $x$ axis \cite{You1}. On the
contrary, the system becomes a spin-spiral phase for $\vert K\vert>K_c$.
Unlike Ising model with XZX-type interactions where the gap-closing
momentum $k_c$ moves in the Brillouin zone along the critical lines
\cite{Niu12}, in our case $k_c$ is suited at Brillouin zone center
$k_c=0$ constrained by the P symmetry.

\begin{figure}[t!]
  \includegraphics[width=8cm]{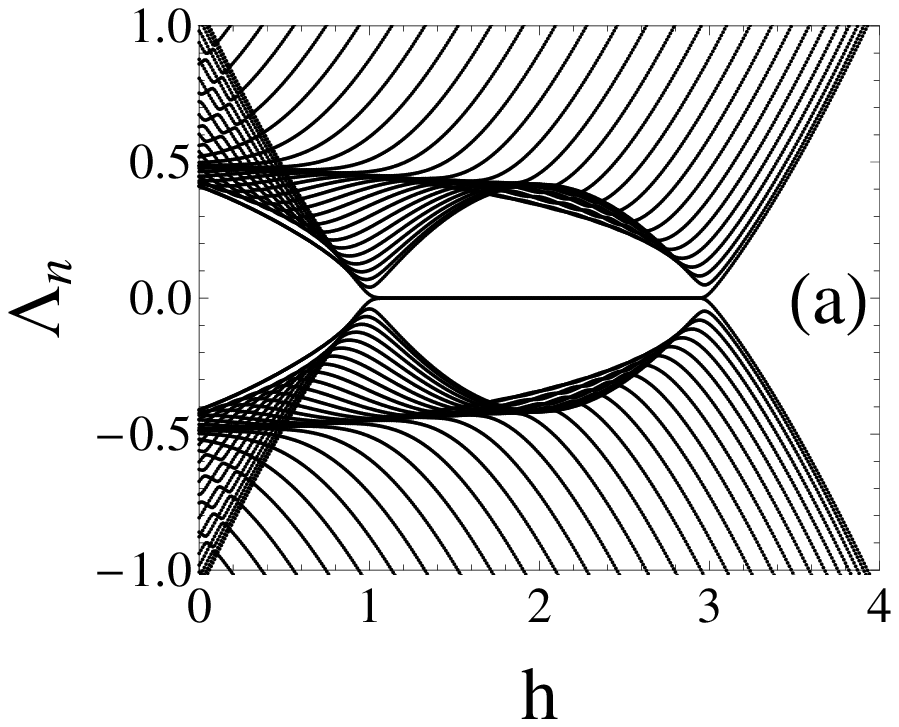}
   \includegraphics[width=8cm]{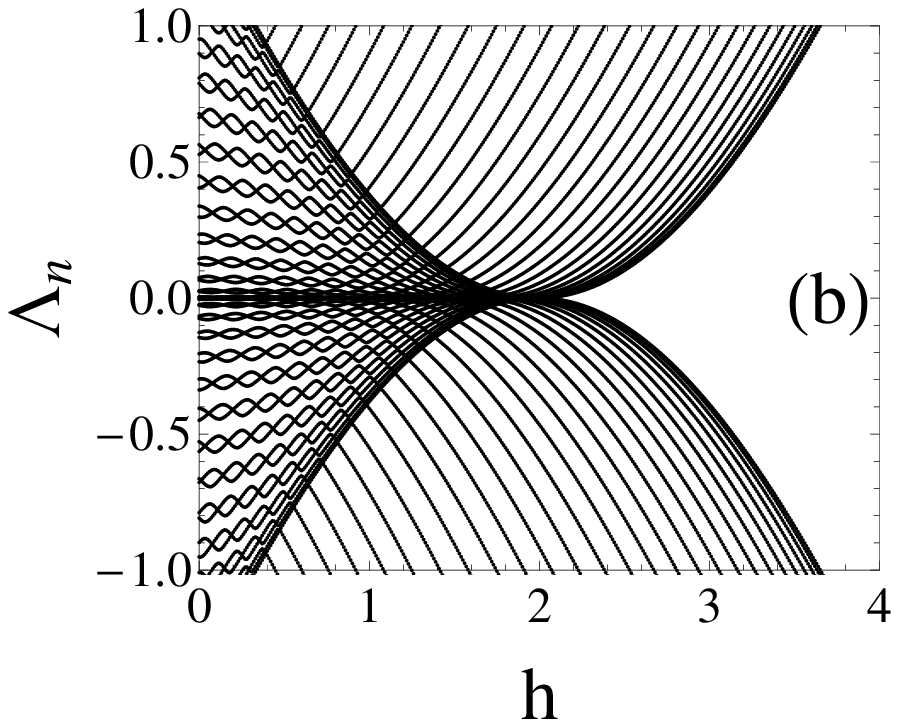}
\caption{ Energy spectrum for the GCM with parameters:
(a) $\theta=\pi/3$, and
(b) $\theta=\pi/2$
under the OBC.
Parameters are as follows: $J_o=1$, $J_e=4$, $K=2$ and $N=80$.}
\label{Fig:Spe_OBC}
\end{figure}

It is clear that the critical lines $K=-\sqrt{J_o J_e}\cos\theta$ and
$K=\sqrt{J_o J_e}\cos\theta$ will get closer as $\theta$ approaches
$\pi/2$. At $\theta=\pi/2$, the 1D GCM Eq. (\ref{model}) describes a
competition between two pseudospin components, $\{\sigma^x_i,\sigma^y_i\}$,
and has the highest possible frustration of interactions.
The mixed terms $\propto\sigma^x_i\sigma^y_{i+1}$ can be eliminated by
writing this model in the form of the GCM with rotated pseudospin
components, where the rotation by angle $\theta=\pi/4$ anticlockwise
with respect to the $z$ axis in the pseudospin space is made, i.e.,
$\bar{\sigma}_{i}^x$ $\equiv$ $ (\sigma_{i}^x + \sigma_{i}^y)/\sqrt{2}$,
$\bar{\sigma}_{i}^y $ $\equiv$ $( \sigma_{i}^y-\sigma_{i}^x)/\sqrt{2}$,
and one finds
\begin{eqnarray}
\cal{\bar{H}}&=& \sum_{i=1}^{N^\prime}
\left(J_{o}\bar{\sigma}_{2i-1}^x \bar{\sigma}_{2i}^x
+J_{e}\bar{\sigma}_{2i}^y \bar{\sigma}_{2i+1}^y \right)\nonumber \\
&+&\sum_{i=1}^{N} 
K(\bar{\sigma}^x_{i-1} \sigma^z_{i}\bar{\sigma}^x_{i+1}
+\bar{\sigma}^y_{i-1} \sigma^z_{i}\bar{\sigma}^y_{i+1}).
\end{eqnarray}
By performing a similar analytical process as Eq. (\ref{FT2}), we can
diagonalize the rotated Hamiltonian in the form of
\begin{eqnarray}
\cal{\bar{H}}&=&
   \sum_{k}
\bar{\Gamma}_k^{\dagger}
\hat{\bar{M}}_k
\bar{\Gamma}_k,
\end{eqnarray}
where
\begin{eqnarray}
\hat{\bar{M}}_k
&=&\frac{1}{2} \left(\begin{array}{cccc}
 \bar{F}_k &  \bar{S}_k &  0   &  \bar{T}_k  \\
\bar{S}_k^* &  \bar{F}_k & -\bar{T}_{-k}    & 0     \\
0 &  -\bar{T}_{-k}^*    & -\bar{F}_k & -\bar{S}_k\\
 \bar{T}_k^*   & 0   &  -\bar{S}_k^* &-\bar{F}_k
\end{array}\right),
\label{HamiltonianMatrix3}
\end{eqnarray}
with
\begin{eqnarray}
\label{Tk2}
 \bar{T}_k&=& J_o+J_e e^{ik},  \nonumber \\
\label{Sk2}
 \bar{S}_k&=& J_o+J_e e^{ik},    \nonumber \\
\label{Fk2}
\bar{F}_k &=& 2 K \cos k, \quad
\end{eqnarray}
and $\bar{\Gamma}_k^{\dagger}=(\bar{a}_{k}^{\dagger},
\bar{b}_k^{\dagger},\bar{a}_{-k},\bar{b}_{-k})$.
Then we have
\begin{eqnarray}
\varepsilon_{k,1(2)}=
\sqrt{\vert\bar{T}_k\vert^2+\vert\bar{F}_k\vert^2}\pm\vert\bar{T}_k\vert.
\end{eqnarray}
It is evident that there is a zero-energy flat band for $K=0$ which is
susceptible to residual interactions. We note that $\bar{F}_k$
(\ref{Fk2}) is vanishing at commensurate momenta $k=\pm\pi/2$.
Therefore, the system turns to be gapless, as recognized in Fig.
\ref{Fig:PBCspectrum}(b).

\section{GENERALIZED COMPASS MODEL IN A~HOMOGENOUS MAGNETIC FIELD}
\label{sec:field}

\begin{figure}[t!]
 \includegraphics[width=9cm]{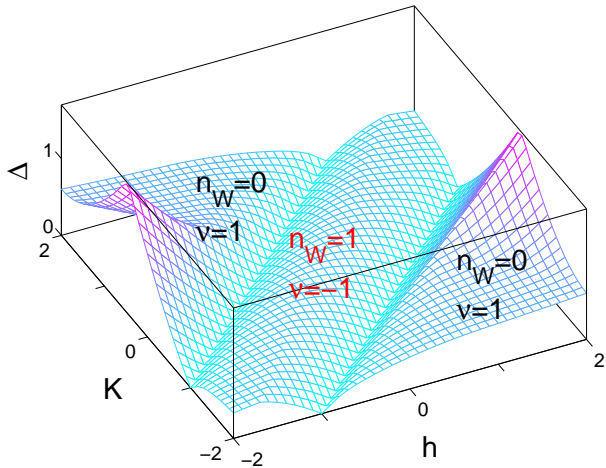}
\caption{The excitation gap $\Delta$ as functions of three-site
interactions $\propto K$ and magnetic field $h$. Here $n_W$ denotes
the number of Majorana zero modes and $\nu$ characterize the Pfaffian
topological invariant.
Parameters are as follows: $J_o=1$, $J_e=4$, $\theta=\pi/3$.}
\label{Fig2:gap2}
\end{figure}

Here we study the effect of a homogenous magnetic field. We consider
the case where the magnetic field is oriented perpendicular to the
easy plane of the spins, i.e., $\vec{h}_o$=$\vec{h}_e$=$h\hat{z}$.
$h$ is the magnitude of the transverse external field, which
contains the $g$-factor $g$ and the Bohr magneton $\mu_{\rm B}$.

\begin{figure}[b!]
 \includegraphics[width=8cm]{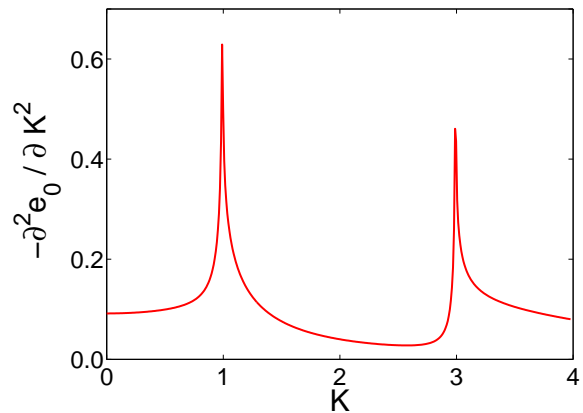}
\caption{The second derivative of the ground-state energy density,
$-(\partial^2 e_0/\partial K^2)$,
for $h=2$. Parameters are as follows: $J_o=1$, $J_e=4$, $\theta=\pi/3$.}
  \label{D2E0}
\end{figure}

The magnetic field does not spoil the zero-energy edge states at
$\theta=\pi/3$, as shown in Fig. \ref{Fig:Spe_OBC}(a). The inclusion
of homogenous magnetic fields replaces $F_k$ in Eq.
(\ref{HamiltonianMatrix2}) with $F_k\to F_k^{'}=2K\cos k-2h$.
The gap as a function of $K$ and $h$ is shown in Fig. \ref{Fig2:gap2}.
The critical lines are pinpointed at $K-h=\pm 1$ for $J_o=1$, $J_e=4$,
$\theta=\pi/3$, as depicted in Fig. \ref{Fig2:gap2}. It is easy to see
that the critical lines found at $h=\pm 1$ in the absence of $K$ are
moved to $K-h=\pm 1$ when the additional (XZX$+$YZY)-type interaction
emerges. In the phase diagram of Fig. \ref{Fig2:gap2} at least three
phases can be specified:
two $z$-axis polarized phases for positive (negative) $h$, and
a canted N\'{e}el (CN) phase for moderate $h$. Such QPTs are of second
order since the second derivative of the ground-state energy density
$e_0$ exhibits divergence, as shown in Fig. \ref{D2E0}.

In the limit of large $h$, the system stays in a polarized state with
$\langle \sigma_i^x \sigma_{i+1}^x \rangle\to 0$,
$\langle \sigma_i^y \sigma_{i+1}^y \rangle\to 0$, and
$\langle \sigma_i^z \sigma_{i+1}^z \rangle\to~1$. In contrast, in the
limit of large $K$ all the nearest neighbor spin correlations vanish,
corresponding to a spiral spin state.
According to the phase diagram of Ising model with (XZX$+$YZY)
interactions given in Ref. \cite{Lian11}, the existence of an additional
phase IV was suggested. However, such a phase is not confirmed in our
investigation, and we believe that a crossover from the spin-spiral
state to the spin-polarized state takes place instead.

\begin{figure}[b!]
\includegraphics[width=8cm]{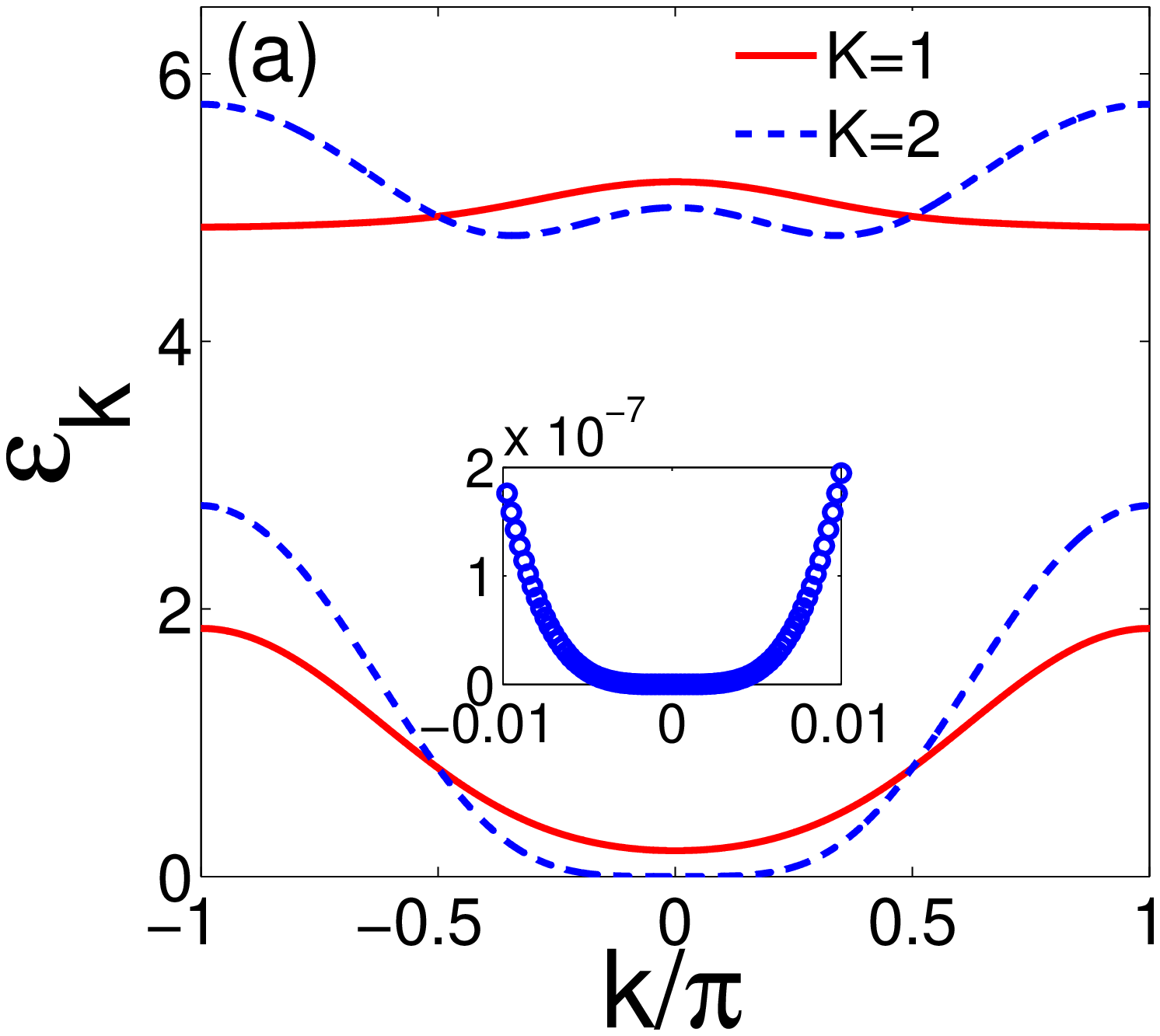}
\includegraphics[width=8cm]{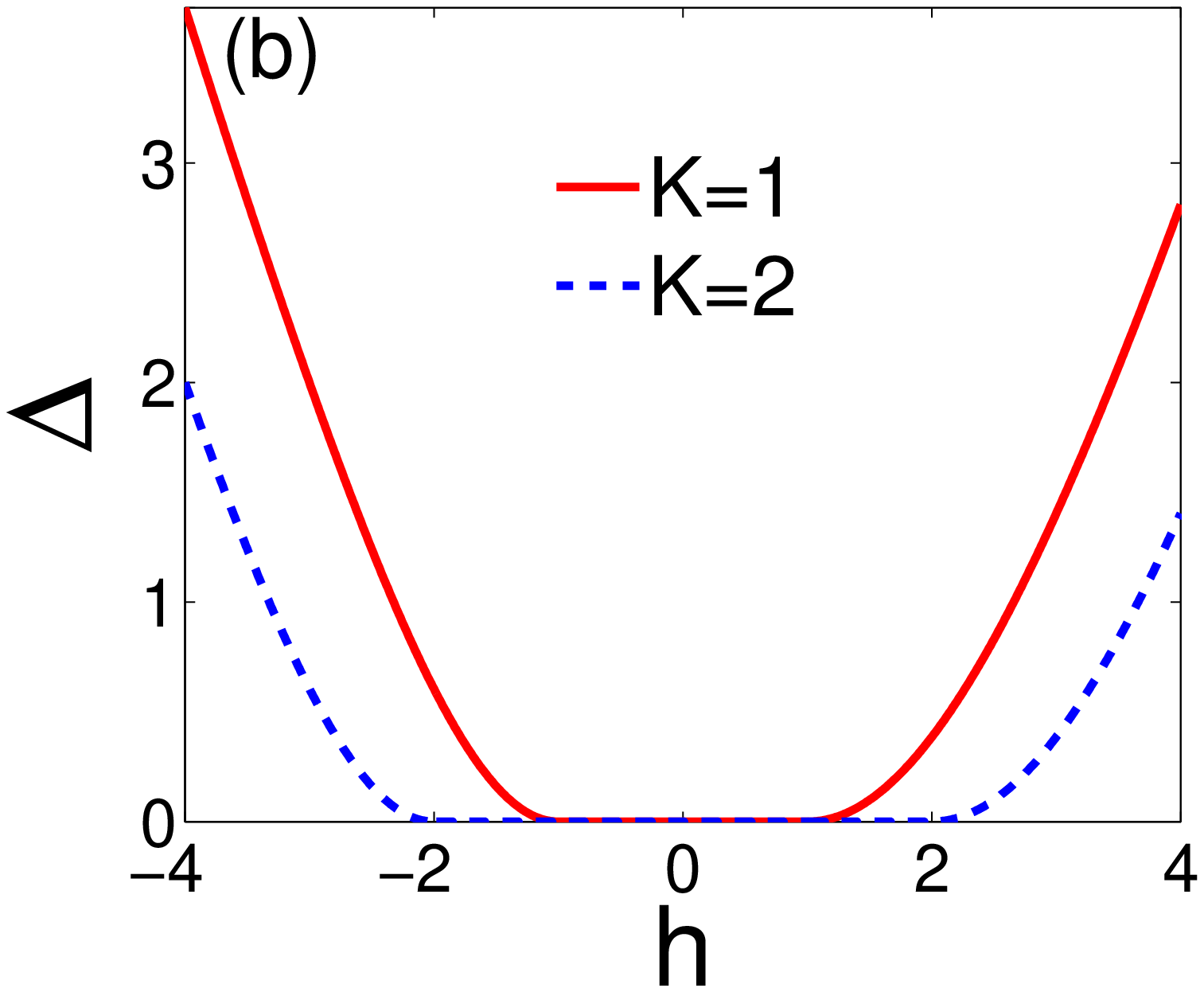}
\caption{
(a) The dispersion relations for different $K$ with $h=2$;
(b) The excitation gap $\Delta$ as functions of $h$ for different
$K=1, 2$. Inset in (a) shows the amplification of lower energies
$\varepsilon_{k,1}$ near Fermi surface at $K=2$.
Parameters are as follows: $J_o=1$, $J_e=4$, $\theta=\pi/2$.}
 \label{Fig2:gap1}
\end{figure}

The critical
behavior is determined by those low-energy states near the critical
mode ($k \sim k_c$). As $h$ approaches $h_c$, the gap vanishes as
$\Delta\sim(h-h_c)^{\nu_h z}$, where $\nu_h$ and $z$ are the
correlation length and dynamic exponents, respectively.
The gap near criticality is
\begin{equation}
\label{Delta}
\Delta \simeq \frac{4 \sqrt{J_o J_e}\cos \theta }{ J_{\rm o}+J_{\rm e}}
\,\vert h-h_c\vert,
\end{equation}
and one finds the critical exponent $\nu_h z=1$.
Since the size dependence of the gap, $\Delta\sim L^{-z}$, defines the
dynamic exponent $z$, we expand the gap around the critical line $h_c$
from threshold critical mode $k_c$, i.e., at $\vert k-k_c\vert \ll 1$,
\begin{eqnarray}
\varepsilon_k &\sim&  \frac{\sqrt{\varsigma_k^2-\tau_k}}{\sqrt{2\varsigma_k}}
\sim \frac{2 J_o J_e \cos^2 \theta}{  J_o +J_e }\,\vert k \vert.
\end{eqnarray}
The dynamic critical exponent $z$ relates the scaling of energy
(or time) scales to length scales. The relativistic spectra at
$k_{\rm c}$ imply a dynamical exponent $z=1$ [for $\theta \neq \pi/2$
in inset of Fig. \ref{Fig:PBCspectrum}(a)] and the Fermi velocity is
independent of $h$ and $K$. The correlation-length exponent $\nu_h=1$
here confirms that 1D GCM belongs to the same universality as the 1D
Ising model under the transverse field.

For $\theta =\pi/2$, a finite magnetic field will modify the energy
spectra through $\bar{F}_k$ $\to$ $\bar{F}_k^{'}=2 K \cos k-2h$ in Eq.
(\ref{HamiltonianMatrix3}), as is uncovered in Fig. \ref{Fig2:gap1}(a).
To this end, $\bar{F}_k^{'}$ can be zero when $\vert h/K \vert \le 1$,
and this causes a closure of the gap at an incommensurate momentum
$k_c = \arccos(h/K)$. Therefore, the system remains gapless as long
as $\vert h/K \vert \le 1$, as evidenced in Fig. \ref{Fig2:gap1}(b).
There is no spontaneous symmetry breaking in this spin-liquid phase
across the quantum critical point (QCP), since the ground-state energy
density,
\begin{equation}
e_0=-\frac{1}{N}\sum_{k} \sqrt{J_o^2+J_e^2+2J_o J_e\cos k + 4 (K \cos k-h)^2 },
\end{equation}
is infinitely differentiable during this transition. The phase
transition is a Berezinskii-Kosterlitz-Thouless (BKT) transition.
In the spin-liquid phase, one can find the spectra vanish at $k_c$,
 \begin{eqnarray}
\varepsilon_k \sim
\frac{2(K^2-h^2)}{ \sqrt{J_o^2+J_e^2+2J_o J_e h/K  } }\,(k-k_c)^2.
\label{quadratic}
\end{eqnarray}
Such quadratic dispersion (\ref{quadratic}) corresponds to a dynamical
exponent $z=2$ [see inset of Fig. \ref{Fig:PBCspectrum}(b)].
While expanding the gap around the QCP from upper threshold
one finds the excitations follow a power-law dependence on $k$,
\begin{eqnarray}
\varepsilon_k
\sim
  \frac{h^2}{  2(J_o +J_e) }\,k^4.
\end{eqnarray}
We confirm the dispersion of fermions is described by a biquadratic
parabola [see inset of Fig. \ref{Fig2:gap1}(b)], and the momentum
dependence of the charge excitations suggests a large dynamical
exponent $z=4$. This leads to a higher density of states above the gap
$D(\varepsilon)\sim\varepsilon^{-3/4}$ than for the standard 1D van
Hove singularity with $D(\varepsilon)\sim \varepsilon^{-1/2}$
(here $\varepsilon$ is the energy measured from the band edge).
Those
low-energy states in the gapless regime near the critical modes
determine the critical behavior. Both the low-temperature entropy
${\cal S}$ and the specific heat $C_V$ present a power-law dependence
on temperature as $T^{d/z}$ (here the spatial dimension $d$ is 1),
which can be readily measured in experiments \cite{Dender97,Kono15,Liang15}.
Meanwhile, the gap near criticality is
\begin{equation}
\label{Delta}
\Delta \simeq \frac{2}{ J_{\rm o}+J_{\rm e} }\, \vert h-h_c\vert^2,
\end{equation}
and one finds the critical exponent $\nu_h z=2$. The outcome $\nu_h=1/2$
for the 1D compass model is different from other points
\cite{Sun09a,Motamedifar13} obtained from scaling of fidelity
susceptibility, which is discussed in Sec. \ref{sec:fs}. The unusual
behavior which takes place due to the multicriticality of such QPTs has
been recognized \cite{Eri09}. Such anomalous feature, such as a flat
dispersion like $k^4$ resembles QPTs between the Mott insulator and
metal in 2D square lattice by controlling the filling \cite{Imada1999}.
Such a new universality class may be characterized
by an emergent super-symmetry at a multicritical point \cite{Huijse15}.

\section{Quantum information theoretical measures}
\label{sec:qi}

Interdisciplinary studies have harvested rich but rather mixed research
findings in the past decades. A blooming topic is the characterization
of QPTs in terms of the ideas from the field of quantum information in
recent years. Different from traditional descriptions of phase
transitions in the theory of condensed matter, the local order parameters,
key ingredients of Ginzburg-Landau-Wilson paradigm, are not necessary in
such a formalism. Instead, quantum information approaches tend to
capture the nonlocal information and universal properties near
criticality despite the great diversity of the nature of miscellaneous
phases.

It should be emphasized that the entanglement entropy \cite{Osterloh02,Gu04}
and the fidelity susceptibility are frequently considered. As a new
perspective of the phase transitions and the associated universality,
they have proven to be useful measures. In this respect, when a quantum
system moves across a QPT separating two fundamentally different
ground states by varying external control parameters, physical
observables often exhibit singular behavior which is ascribed to the
gapless excitation and divergence of correlation length at the QCPs.
Frequently this picture can be visualized when there are symmetry
breaking states on either side or both sides of a QPT.
However, a topological phase transition follows from a change of
topological index of the ground state and the topological phase of
matter is not related to the spontaneous symmetry breaking. Therefore,
a local order parameter has no scope for its ability to sense the
topological QPT.

\subsection{Fidelity susceptibility}
\label{sec:fs}

\begin{figure}[t!]
\includegraphics[width=8cm]{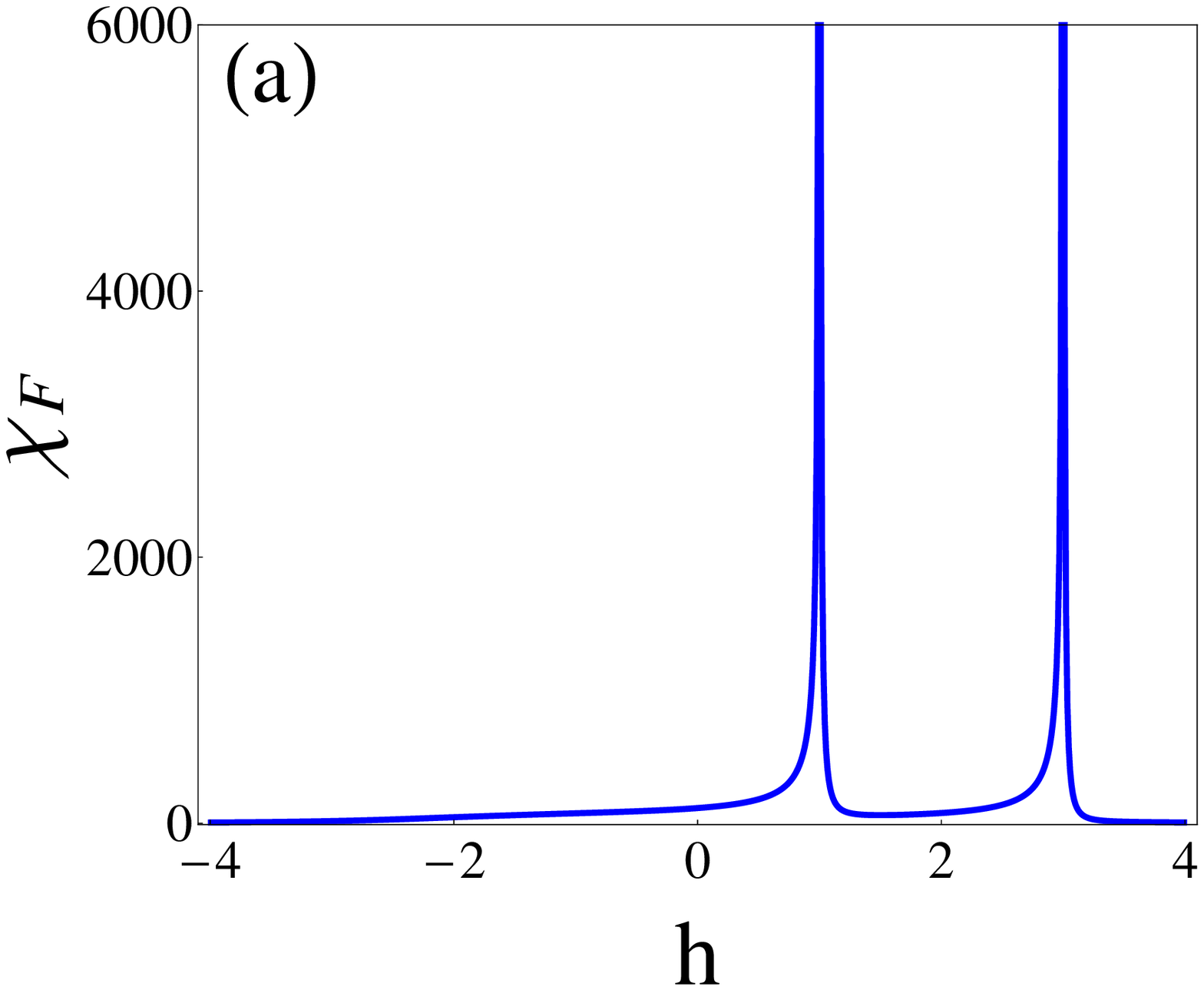}
\includegraphics[width=8cm]{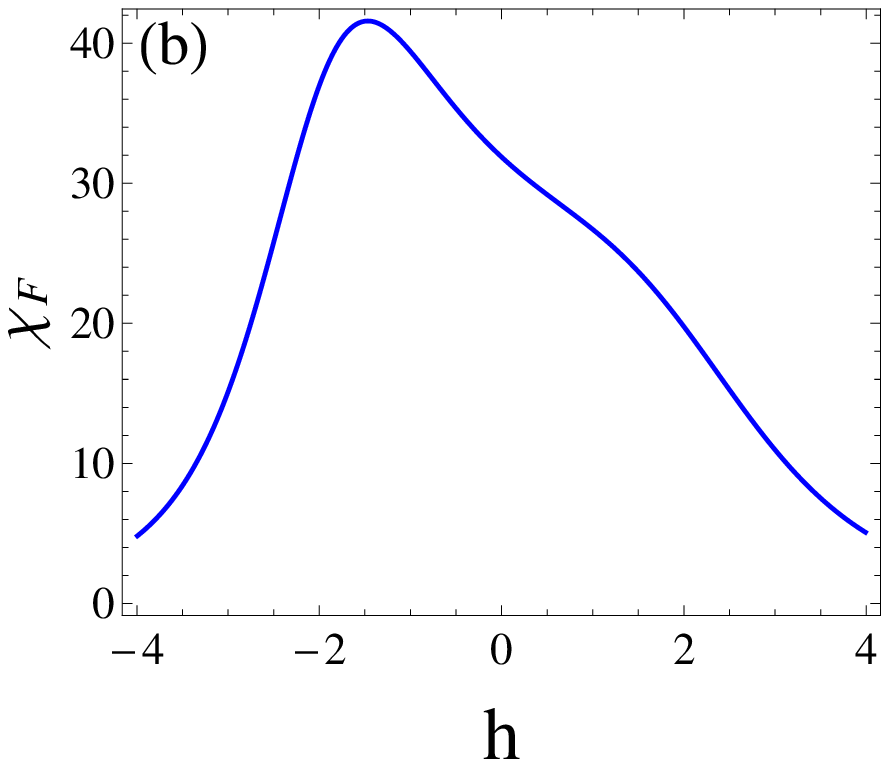}
\caption{The fidelity as a function of $h$ for the GCM with three-site
interactions (\ref{model}) for:
(a) $\theta=\pi/3$ and
(b) $\pi/2$.
Parameters: $J_o=1$, $J_e=4$, $K=2$, $N=1000$.}
 \label{Fid:1}
\end{figure}

The fidelity susceptibility is a general probe of phase transition
which originates from Anderson's orthogonality catastrophe.
By definition, quantum fidelity of a many-body Hamiltonian
$\hat{H}(\lambda)=H_0 + \lambda H_I$ is \cite{Zanardi07}
\begin{eqnarray}
F(\lambda_0,\lambda_1)=
\vert\langle\Psi_0(\lambda_0)\vert\Psi_0(\lambda_1)\rangle\vert,
\end{eqnarray}
where $\vert \Psi_0 \rangle$ is the ground state, $\lambda_0$ and
$\lambda_1$ specify two points in the parameter space of driving
parameter $\lambda$. In this respect, fidelity susceptibility is
defined as first nonzero order of the Taylor expansion of the overlap
function $F(\lambda,\lambda+\delta \lambda)$, given by
\cite{You07,Venuti07}
 \begin{eqnarray}
\chi_F= \lim_{\delta \lambda \to 0}
\frac{-2 \ln F(\lambda,\lambda+\delta \lambda)}{(\delta \lambda)^2}.
\label{chiform}
\end{eqnarray}
The concept of quantum fidelity susceptibility has been recognized as
a versatile indicator in identifying QCPs and universality class by the
finite-size scaling behavior \cite{Gu10}. Interestingly, a holographic
description for the fidelity susceptibility in conformal field theories
is a volume of maximal time slice in an anti-de Sitter space-time when
the perturbation is exactly marginal \cite{Miyaji15}. However, the
application of the fidelity susceptibility to detect a BKT transition
is controversial: On the one hand, some investigations are in favor that
the fidelity susceptibility is able to discriminate the critical lines
of BKT transitions with a logarithmic divergence \cite{Yang07,Wang10},
while on the other hand, some disprove it \cite{Sun15,You15}. This
shows that indeed an in-depth understanding of the underlying physics
is still missing.

We add to this discussion and present the fidelity susceptibility for
$\theta=\pi/3$ and $\theta=\pi/2$ with $K=1$; more details of the
calculation can be found in Appendix \ref{sec:appa}. The fidelity
susceptibility for $\theta=\pi/3$ detects the second-order QPTs,
seen Fig. \ref{Fid:1}(a), while such a transition is absent for
$\theta=\pi/2$ shown in Fig. \ref{Fid:1}(b). Our findings suggest that
the fidelity susceptibility does not diverge at BKT-type QPTs in one
spatial dimension.

\subsection{Coherence susceptibility}
\label{sec:coh}

In the representation spanned by the two-qubit product states we employ
the following basis,
\begin{equation}
\{\vert 0\rangle_i\otimes\vert 0\rangle_j,
\vert 0\rangle_i\otimes\vert 1\rangle_j,
\vert 1\rangle_i\otimes\vert 0\rangle_j,
\vert 1\rangle_i\otimes\vert 1\rangle_j\},
\end{equation}
where $\vert 0\rangle$ ($\vert 1\rangle$) denotes spin up (down) state,
the two-site density matrix can be expressed as,
\begin{equation}
\rho_{ij}=\frac{1}{4}\sum_{a,a'=0}^3\langle
\sigma_i^{a}\sigma_j^{a'}\rangle
   \sigma_i^{a} \sigma_j^{a'},
\end{equation}
where $\sigma_i^{a}$ are Pauli matrices $\sigma_i^{x}$, $\sigma_i^{y}$,
and $\sigma_i^{z}$ for $a=1,2,3$, and a $2\times 2$ unit matrix for
$a=0$. The Hamiltonian has $\mathbb{Z}_2$ symmetry, namely, the
invariance under parity transformation $P=\otimes_{i}\sigma_{i}^z$, and
then correlation functions such as $\langle\sigma_i^a\sigma_j^b\rangle$
($a=x,y$ and $b=0,z$) vanish simultaneously. Usually people believe that
 $\langle \sigma_i^{x}\sigma_j^y\rangle$
($\langle \sigma_i^{y}\sigma_j^x\rangle$) vanishes due to the imaginary
character of $\sigma_j^y$ ($\sigma_i^y$). Here we disprove this argument
in our model due to its complex nature of Hamiltonian (\ref{model}) in
Appendix \ref{sec:appb}. Also, be aware that the relations between
correlations where
\begin{equation}
\langle \sigma_0^z\sigma_r^z \rangle=\langle\sigma_0^z\rangle
\langle\sigma_r^z\rangle - G_r G_{-r}
\end{equation}
is not always valid for a complex Hamiltonian [see the definition of
$G_r$ in Eq.(\ref{Gr})]. A number of results have
been focused on translation-invariant systems and almost exclusively
correspond to reflection-symmetric systems, despite the fact that models
violating reflection invariance play a prominent role in many-body
theory, e.g., in describing interactions of DM
interactions or three-site (XZY$-$YZX)-type interactions \cite{Liu11}.

Therefore, the two-qubit density matrix reduces to an X-state,
\begin{equation}
\rho_{ij}=\left(
\begin{array}{cccc}
u^{+} & 0 & 0 & z_{1} \\
0 & w_{1} & z_{2} & 0 \\
0 & z_{2}^{*} & w_{2} & 0 \\
z_{1}^{*} & 0 & 0 & u^{-}%
\end{array}%
\right),  \label{eq:2DXXZ_RDM}
\end{equation}%
with
\begin{eqnarray}
u^{\pm }\!&=&\frac{1}{4}(1\pm  2\langle {\sigma_{i}^{z}}\rangle
+\langle {\sigma _{i}^{z}\sigma _{j}^{z}}\rangle ),
\label{upm} \\
z_{1}\!&=&\frac{1}{4}(\langle \sigma _{i}^{x}\sigma _{j
}^{x}\rangle - \langle \sigma _{i}^{y}\sigma _{j
}^{y}\rangle -i \langle \sigma _{i}^{x}\sigma _{j
}^{y}\rangle - i \langle \sigma _{i}^{y}\sigma _{j
}^{x}\rangle),  \label{z1} \\
z_{2}\!&=&\frac{1}{4}(\langle \sigma _{i}^{x}\sigma _{j
}^{x}\rangle + \langle \sigma _{i}^{y}\sigma _{j
}^{y}\rangle+i \langle \sigma _{i}^{x}\sigma _{j
}^{y}\rangle - i \langle \sigma _{i}^{y}\sigma _{j
}^{x}\rangle ),  \label{z2} \\
\omega _{1}\!&=&\omega _{2}=\frac{1}{4}(1-\langle \sigma _{i
}^{z}\sigma _{j}^{z}\rangle ).  \label{omega1}
\end{eqnarray}%
The density matrix of a single qubit is easily obtained by a partial
trace over one of the two qubits,
\begin{equation}
\rho_{i}=\left(
\begin{array}{c c}
\frac{1}{2}(1+\langle {\sigma_{i}^{z}}\rangle) & 0  \\
0 & \frac{1}{2} (1- \langle {\sigma_{i}^{z}}\rangle)
\end{array}%
\right).
\label{eq:2DXXZ_RDM}
\end{equation}
One easily finds that
\begin{eqnarray}
S(\rho_{i})&=&S(\rho_{j})=-\sum_{m=0}^1\{ [1+(-1)^m \langle
\sigma_{i}^z \rangle ]/2\}\nonumber \\
&\times&\log_2\{ [1+(-1)^m \langle \sigma_{i}^z
\rangle ]/2\},
\end{eqnarray}
and
\begin{eqnarray}
S(\rho_{ij})=-\sum_{m=0}^1 \xi_m \log_2 \xi_m-\sum_{n=0}^1\xi_n\log_2 \xi_n,
\end{eqnarray}
where
\begin{widetext}
 \begin{eqnarray}
 \xi_m &=&  \frac{1}{4}\left[1+ \langle
\sigma_{i}^z \sigma_{j}^z \rangle  +  (-1)^m
\sqrt{(\langle\sigma_{i}^x \sigma_{j}^x\rangle-\langle\sigma_{i}^y
\sigma_{j}^y\rangle)^2+ (\langle\sigma_{i}^x \sigma_{j}^y\rangle+\langle\sigma_{i}^y
\sigma_{j}^x\rangle)^2 + 4 \langle \sigma_{i}^z \rangle^2 }\right] ,
\\
\xi_n &=& \frac{1}{4}\left[1 - \langle \sigma_{i}^z \sigma_{j}^z \rangle +
(-1)^n\sqrt{(\langle\sigma_{i}^x \sigma_{j}^x\rangle+\langle\sigma_{i}^y
\sigma_{j}^y\rangle)^2 + (\langle\sigma_{i}^x \sigma_{j}^y\rangle-\langle\sigma_{i}^y
\sigma_{j}^x\rangle)^2} \right] .
\end{eqnarray}
\end{widetext}

\begin{figure}[b!]
 \includegraphics[width=\columnwidth]{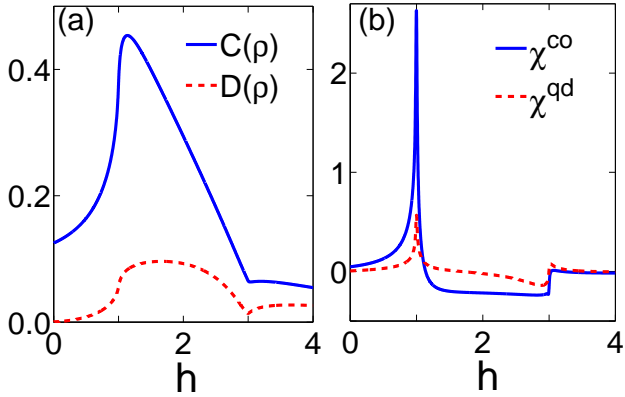}
\caption{Characterization of the ground state for the model Eq.
(\ref{model}) at $\theta=\pi/3$ for increasing magnetic field $h$:
(a) relative entropy $C(\rho)$ and quantum discord $D(\rho)$, and
(b) coherence susceptibility $\chi^{co}$ and discord susceptibility
$\chi^{qd}$. Other parameters: $J_o=1$, $J_e=4$, $K=2$ for $N=200$. }
 \label{copi3}
\end{figure}

A simplified form of relative entropy has been proven as a valid
measure of coherence for a given basis:
\begin{eqnarray}
C({\rho_{ij}}) &=& S({\rho}_{diag}) - S({\rho_{ij}}),
\label{eq:measure}
\end{eqnarray}
where $S(\bullet)$ stands for the von Neumann entropy of $\bullet$ and
${\rho}_{diag}$ is obtained from ${\rho}$ by removing all its
off-diagonal entries. The non-analyticity of the ground state at
QCPs can be characterized by the singularity of the
coherence susceptibility \cite{Chen16}, which is defined as
\begin{eqnarray}
\chi^{co} \equiv \partial C(\rho)/\partial \lambda.
\end{eqnarray}
Here, $\rho$ stands either for the density operator of the whole system
or for the reduced density operator of a subsystem.

\begin{figure}[b!]
\includegraphics[width=\columnwidth]{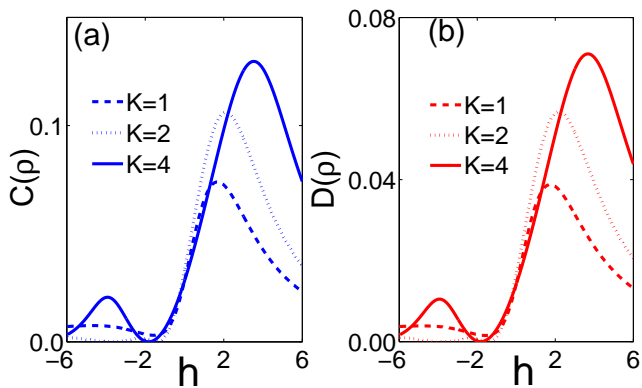}
\caption{Characterization of the ground state for the model Eq.
(\ref{model}) at $\theta=\pi/2$ for increasing magnetic field $h$ and
for selected values of $K$:
(a) relative entropy $C(\rho)$ and
(b) quantum discord $D(\rho)$.
Other parameters: $J_o=1$ and $J_e=4$ for $N=200$.}
\label{Correlation_theta=pi2}
\end{figure}

It was interesting to note that quantum discord, in contrast to
entanglement, is able to signal the BKT-type QPTs
\cite{Dil08,Sarandy09,Werlang10}. The quantum discord was introduced to
quantify non-classical correlations beyond entanglement paradigm in
quantum states and thus was given by the difference of the mutual
information $I(\rho)$ and the classical correlation $J(\rho)$
\cite{Ollivier01},
\begin{eqnarray}
D(\rho)=I(\rho)-J(\rho).
\end{eqnarray}
Similarly we can define discord susceptibility,
\begin{eqnarray}
\chi^{qd} \equiv \partial D(\rho)/\partial \lambda.
 \end{eqnarray}
The relative entropy and quantum discord as functions of $h$ at $K=2$
are plotted in Fig. \ref{copi3}. We find that both quantities share
similar trends and there are sharp changes across the QPTs. The peaks
of their susceptibilities at $h=1$ and the step-like behavior at $h=3$
indicate the QCPs.

We emphasize that quantum correlations for $\theta=\pi/2$ revealed by
the relative entropy and the quantum discord exhibit distinct behavior
from the case with $\theta=\pi/3$, as seen in Fig.
\ref{Correlation_theta=pi2}. Both quantities show their local maxima
close to $h=\pm K$. However, we find that these indictors behave in a
more distinct way in the regime of large $K$. For small $K$ two local
maxima affect each other and move the positions of maxima from the true
QCPs. In addition, we find that concurrence, another measure of
entanglement \cite{Wootters98}, and von Neumann entropy display similar
behaviors with fidelity susceptibility (not shown).

\section{Conclusion}
\label{sec:summa}

In the paper we analyze quantum phase transitions in a class of
the one-dimensional compass models with an (XZX$+$YZY)-type of
three-site interactions. We present the exact solution by means of
Jordan-Wigner transformation, and study the fermionic spectra,
excitation gap, critical exponents, and established the phase diagram.
For general titling angle $\theta$, the three-site (XZX$+$YZY)
interactions renormalize the effect of magnetic field and thus a
nontrivial magnetoelectric effect can be expected. In the canted
N\'{e}el phase (weak-coupling BCS regime in spinless fermions), it
exhibits a pair of zero-energy Majorana modes at each end of the open
chain, and it is also characterized with a Pfaffian topological
invariant $\nu=-1$ with periodic boundary condition.

In the compass
limit the competition between the three-site (XZX$+$YZY) interactions
and the magnetic field drives the system into a gapless phase through
a Berezinskii-Kosterlitz-Thouless transition. The dynamic exponent is
a measure for characterizing the coherence of the system and it is
found to be $z=4$ across the quantum critical points. Thus, coherence
is very sensitive to whether the system is at the compass limit, i.e.,
at the angle $\theta=\pi/2$ which is more incoherent than the other
cases. It has been shown that $z$ can be extracted from the measurement
of the low-temperature specific heat and entropy in the
Tomonaga-Luttinger-liquid phase.

To complete the analytic approach, we present a study of diverse
measures of quantum correlations including fidelity susceptibility, von
Neumann entropy, relative entropy, coherence susceptibility, pairwise
concurrence and quantum discord in the generalized compass chain with
three-site (XZX$+$YZY) interactions.
Analytical expressions are obtained from the spin-spin correlation
functions. We show that all these measures can be useful to detect
the second-order transition, while only the relative entropy and the
quantum discord can signal the Berezinskii-Kosterlitz-Thouless
transition. We note that the one-dimensional compass model with
(XZX$+$YZY)-type interactions can provide an ideal benchmark for other
computational methods to testify the Berezinskii-Kosterlitz-Thouless
quantum phase transition. We also point out that deriving the
correlation functions for the systems with broken reflection symmetry
requires a rather careful and subtle procedure.

\acknowledgments
We thank G.-S. Tian, Dazhi Xu, Yunfeng Cai, Y.-L. Dong and Hua Jiang for
insightful discussions. This work was
supported by the Natural Science
Foundation of Jiangsu Province of China under Grant No. BK20141190
and the NSFC under Grant Nos. 11474211, 11504253, 21473240.
A.~M.~O. kindly acknowledges support by Narodowe Centrum Nauki
(NCN, National Science Center) under Project No.~2012/04/A/ST3/00331.

\appendix

\section{Eigenstates and eigenvalues of generalized compass model}
\label{sec:appa}

By Fourier transforming the GCM Hamiltonian (\ref{HamiltonianMatrix2})
and grouping together terms with $k$ and -$k$, $H$ is transformed into
a sum of commuting Hamiltonians $H_k$ describing a different $k$ mode
each. Then we can obtain the spectrum of the GCM by diagonalizing each
Hamiltonian mode $H_k$ independently.

Formally we write the Hamiltonian mode $H_k$ in the BdG form,
\begin{eqnarray}
\hat{H}_{k} = \Gamma_k^{\dagger}\hat{M}_k\Gamma_k, \label{eq:Mk3}
\end{eqnarray}
and $\Gamma_k^{\dagger}\equiv(a_k^{\dagger},a_{-k}^{},b_k^{\dagger},b_{-k}^{})$.
$\hat{M}_k$ can be diagonalized by a unitary transformation,
\begin{eqnarray}
\hat{H}_k  &=& \Gamma_{k}^{\dagger} U_k^{} U_K^\dagger M^\prime_k
U_K^{} U_k^\dagger \Gamma_{k}^{}  \nonumber \\
&=& \sum_{k}\Gamma_{k}^{\prime\dagger} D_{k}^{} \Gamma_{k}^{\prime},
\end{eqnarray}
where $\Gamma_{k}^{\prime}$=$U_k^\dagger\Gamma_{k}^{}$, i.e.,
the diagonalized form is achieved by a four-dimensional Bogoliubov
transformation which connects the original operators
$\{a_k^{\dagger},a_{-k}^{},b_k^{\dagger},b_{-k}^{}\}$ with two kind of
quasiparticles
$\{\gamma_{k,1}^{\dagger},\gamma_{-k,1}^{},\gamma_{k,2}^{\dagger},\gamma_{-k,2}^{}\}$
as follows,
\begin{eqnarray}
\left(
\begin{array}{c}
\gamma_{k,1}^{\dagger} \\
\gamma_{-k,1}  \\
\gamma_{k,2}^{ \dagger}\\
\gamma_{-k,2}
\end{array}
\right)=\hat{U}_{k} \left(
\begin{array}{c}
a_k^{\dagger}  \\
a_{-k}   \\
b_k^{\dagger}   \\
b_{-k}
\end{array}%
\right), \label{eq:2DXXZ_RDM}
\end{eqnarray}

The obtained four eigenenergies $\{\varepsilon_{k,j}\}$ ($j=1,\cdots,4$)
are the excitations in the artificially enlarged particle-hole space
where the positive (negative) ones denote the electron (hole)
excitations. The ground state corresponds to the state in which all hole
modes are occupied while the electron modes are vacant. The PHS
indicates here that
\begin{eqnarray}
\varepsilon_{k,1 (2)}&=&\sqrt{\varsigma_k+(-1)^j\sqrt{\tau_k }}\; >0, \nonumber\\
\varepsilon_{k,4 (3)}&=&-\varepsilon_{-k,1 (2)} <0. \nonumber
\end{eqnarray}
The diagonal form of the Hamiltonian model,
\begin{eqnarray}
\hat{H}_{k} &=&  \frac{1}{2}\varepsilon_{k,1}\left(
\gamma_{k,1}^{\dagger}\gamma_{k,1} -
\gamma_{-k,1}\gamma_{-k,1}^{\dagger}\right) \nonumber \\
&+&\frac{1}{2}\varepsilon_{k,2}\left(
\gamma_{k,2}^{\dagger}\gamma_{k,2} -
\gamma_{-k,2}\gamma_{-k,2}^{\dagger}\right) \nonumber \\
&=& \sum_{j=1}^{2}\,\varepsilon_{k,j}\left(\gamma_{k,j}^{\dagger}\gamma_{k,j}-\frac12\right).
\label{exc}
\end{eqnarray}

On the other hand, we can use a basis in which the eigenstates of $H_k$
are obtained as linear combinations of even-parity fermion states. Here
we outline the connection between these two approaches. A general
eigenstate is
\begin{eqnarray}
|\psi_{m,k}\rangle\!&=& v_{1}^{m}|0\rangle+v_{2}^{m}~a_{k}^{{\dag}}a_{-k}^{{\dag}}|0\rangle+
v_{3}^{m}~a_{k}^{{\dag}}b_{-k}^{{\dag}}|0\rangle \nonumber \\
&+&\!\!v_{4}^{m}~a_{-k}^{{\dag}}b_{k}^{{\dag}}|0\rangle
+\!v_{5}^{m}~b_{k}^{{\dag}}b_{-k}^{{\dag}}|0\rangle
+\!v_{6}^{m}~a_{k}^{{\dag}}a_{-k}^{{\dag}}b_{k}^{{\dag}}b_{-k}^{{\dag}}|0\rangle,\nonumber \\
\label{comb}
\end{eqnarray}
with $m=1,2,\cdots,6$. In other words, we introduce basis vectors for every $k$,
 \begin{eqnarray}
\vert \varphi_{1,k}\rangle\!&=&\!\vert 0 \rangle, \hskip 1.2cm
\vert \varphi_{2,k}\rangle =a_{k}^{{\dag}}a_{-k}^{{\dag}}|0\rangle, \nonumber \\
\vert \varphi_{3,k}\rangle\!&=&\!a_{k}^{{\dag}}b_{-k}^{{\dag}}|0\rangle, \hskip .3cm
\vert \varphi_{4,k}\rangle = a_{-k}^{{\dag}}b_{k}^{{\dag}}|0\rangle, \nonumber \\
\vert \varphi_{5,k}\rangle\!&=&\!b_{k}^{{\dag}}b_{-k}^{{\dag}}|0\rangle, \hskip .3cm
\vert \varphi_{6,k}\rangle =
a_{k}^{{\dag}}a_{-k}^{{\dag}}b_{k}^{{\dag}}b_{-k}^{{\dag}}|0\rangle.
\end{eqnarray}
The subspace used in Eq. (\ref{comb}) is six-dimensional which is due
to the selected pairs from four modes. In this case,
\begin{eqnarray}
\hat{H}_{k} = \Sigma_k^{\dagger}
\hat{\Xi}_k\Sigma_k,
\end{eqnarray}
where $\hat{\Xi}_k$ can be written explicitly in terms of the matrix
elements of $\hat{M}_k$ in Eq. (\ref{eq:Mk3}):
\begin{widetext}
 \begin{eqnarray}
\hat{\Xi}_k =\left(
               \begin{array}{cccccc}
                 M_{22}+M_{44} & M_{12} & M_{14} & -M_{32} & M_{34} & 0 \\
                 M_{21} & M_{11}+M_{44} & -M_{24} & -M_{31} & 0 & M_{34} \\
                 M_{41} & -M_{42} &  M_{11}+M_{22} & 0 & M_{31} & -M_{32} \\
                 -M_{23} & -M_{13} & 0 & M_{33}+M_{44} & M_{24} & M_{14} \\
                 M_{43} & 0 & M_{13} & M_{42} & M_{22}+M_{33} & M_{12} \\
                 0 & M_{43} & -M_{23} & M_{41} & M_{21} & M_{11}+M_{33} \\
               \end{array}
             \right).
\end{eqnarray}
\end{widetext}

\section{To diagonalize a general Hamiltonian quadratic in fermions }
\label{sec:appb}

The presented analytic method requires diagonalization of a general
quadratic Hamiltonian of the form,
\begin{eqnarray}
H&=& \sum_{i,j}\left\{c_{i}^{\dagger} A_{i j}c_{j}
+\frac{1}{2}\left(c_{i}^{\dagger}B_{i j}c_{j}^{\dagger}
-c_{i}^{}B_{i j}^* c_{j}^{}\right)\right\},
\label{ABHam}
\end{eqnarray}
where $c_i^{}$ and $c_i^\dagger$ are fermion annihilation and creation
operators respectively. For system size $N$, $A$ and $B$ are both
$N\times N$ matrices. The requirement of translation-invariance implies
that $A$ and $B$ are Toeplitz matrices, i.e.,
$A_{i+n,j+n}=A_{i,j}$ and $B_{i+n,j+n}=B_{i,j}$ for any $n\in{\mathbb N}$.
Hermiticity of $H$ implies that $A$ is a (possibly complex) Hermitian
matrix and $B$ is a (possibly complex) antisymmetric matrix.
Finite-ranged interaction means that there exists a positive integer
$l_0$ such that $A_{0,l}=B_{0,l}=0$ if $l\ge l_0$.

Such a spin-chain Hamiltonian is not invariant with respect to the
reflection transformation $R(\sigma^{a}_i)=\sigma^{a}_{-i}$ $(a=x,y,z)$
when $A$ is not a real matrix.
One of the typical quantum spin chains with broken reflection
symmetry is the Ising model with transverse magnetic field and
Dzyaloshinskii-Moriya interaction (in the $z$-direction).
Extra care should be taken that $B$ matrix is not real in these cases.

Normally, people believe that $\langle \sigma_i^{x}\sigma_j^{y}\rangle$
($\langle \sigma_i^{y}\sigma_j^{x}\rangle$) is zero due to the
imaginary character of $\sigma_j^{y}$ ($\sigma_i^{y}$). Here we disprove
this argument in our model due to its complex nature of Hamiltonian.
Also, be aware that the relations between correlations where
\begin{equation}
\langle \sigma_0^z\sigma_r^z \rangle=
\langle \sigma_0^z \rangle \langle \sigma_r^z \rangle-G_r G_{-r},
\end{equation}
is not always valid. In this appendix we will show that this is not
correct. Here
\begin{eqnarray}
G_r=\langle B_0  A_r  \rangle, \label{Gr}
\end{eqnarray}
where
\begin{eqnarray}
A_i\equiv c_{i}^{\dagger}+c_i^{}, \quad B_i\equiv c_{i}^{\dagger}-c_i^{}.
\end{eqnarray}
We can write
\begin{eqnarray}
\sigma_i^{x}&=&A_i \prod_{j=1}^{i-1} A_j B_j, \nonumber \\
\sigma_i^{y}&=& i B_i \prod_{j=1}^{i-1} A_j B_j,\nonumber \\
\sigma_i^{z}&=&  A_i B_i,
\end{eqnarray}
and we have
\begin{eqnarray}
\rho_{i,i+1}^{xx}&=&\langle B_i A_{i+1} \rangle , \nonumber \\
\rho_{i,i+1}^{yy}&=&-\langle A_i B_{i+1} \rangle ,\nonumber \\
\rho_{i,i+1}^{zz}&=&  \langle A_i B_i A_{i+1}B_{i+1} \rangle \nonumber \\
&=& \langle A_i B_{i} \rangle \langle A_{i+1} B_{i+1} \rangle
- \langle A_i B_{i+1} \rangle \langle A_{i+1} B_{i} \rangle \nonumber \\
&-&\langle A_i A_{i+1} \rangle \langle B_{i} B_{i+1} \rangle.
\end{eqnarray}
Generally speaking, they are claimed to obey the algebra irrespective
of detailed eigenspectrum,
\begin{eqnarray}
\{ A_i,A_j \}= 2\delta_{ij},\, \{B_i,B_j \}= -2\delta_{ij},\, \{ A_i,B_j \}=0.
\end{eqnarray}
However, one may be easily verified that it is not true.
Take nearest neighbor sites for example, i.e., $\vert j-i \vert=1$ and one has
\begin{eqnarray}
\langle A_i A_{i+1} \rangle =-i  \langle \sigma_i^y \sigma_{i+1}^x \rangle, \nonumber   \\
\langle B_i B_{i+1} \rangle =-i  \langle \sigma_i^x \sigma_{i+1}^y \rangle,
\end{eqnarray}
In what follows we concentrate on the correlations between the nearest
neighbor spins. By straightforward calculation it is found that the
nearest neighbor spin correlation function has the form
$\langle\sigma_i^x\sigma_{i+1}^y\rangle=
-\langle\sigma_i^y\sigma_{i+1}^x\rangle=i\langle B_i B_{i+1}\rangle$,
so element $z_1$ for the nearest neighbor spins is always a real number
and $z_2$ may be a complex number depending on which phase the system
is in.

Therefore, the two-qubit density matrix reduces to an X-state,
\begin{equation}
\rho_{ij}=\left(
\begin{array}{cccc}
u^{+} & 0 & 0 & z_{1} \\
0 & w^{+} & z_{2} & 0 \\
0 & z_{2}^{*} & w^{-} & 0 \\
z_{1}^{*} & 0 & 0 & u^{-}
\end{array}
\right),
\label{eq:2DXXZ_RDM}
\end{equation}
with
\begin{eqnarray}
u^{\pm }&=&\frac{1}{4}(1\pm\langle{\sigma_{i}^{z}}\rangle
\pm\langle{\sigma_{j}^{z}}\rangle
+\langle {\sigma _{i}^{z}\sigma _{j}^{z}}\rangle ),\nonumber
\label{upm} \\
z_{1}&=&\frac{1}{4}(\langle \sigma _{i}^{x}\sigma _{j}^{x}\rangle
- \langle \sigma _{i}^{y}\sigma _{j}^{y}\rangle
-i \langle \sigma _{i}^{x}\sigma _{j}^{y}\rangle
- i \langle \sigma _{i}^{y}\sigma _{j}^{x}\rangle),  \nonumber
\label{z1} \\
z_{2}&=&\frac{1}{4}(\langle \sigma _{i}^{x}\sigma _{j}^{x}\rangle
+ \langle \sigma _{i}^{y}\sigma _{j}^{y}\rangle
+i \langle \sigma _{i}^{x}\sigma _{j}^{y}\rangle
- i \langle \sigma _{i}^{y}\sigma _{j}^{x}\rangle ), \nonumber
\label{z2} \\
\omega^{\pm}&=&\frac{1}{4}(1\pm \langle{\sigma_{i}^{z}}\rangle
\mp\langle {\sigma_{j}^{z}}\rangle
-\langle\sigma _{i}^{z}\sigma_{j}^{z}\rangle ).  \nonumber
\label{omega1}
\end{eqnarray}%
When the system is translation invariant, we obtain
$\langle{\sigma_{i}^{z}}\rangle=\langle{\sigma_{j}^{z}}\rangle$
($\forall i,j$) such that $\omega^+=\omega^-$.
This missing of terms like $\langle\sigma _{i}^{x}\sigma_{j}^{y}\rangle$,
$\langle\sigma_{i}^{y}\sigma _{j}^{x}\rangle$ commonly exist in Ref.
\cite{Liu11} or a negligence taking
$\langle c_i^\dagger c_{i+1}^\dagger \rangle= 0$ for granted in
calculations of $\langle \sigma _{i}^{z}\sigma _{j}^{z}\rangle$
\cite{Lei15}.

Finally, by a numerical calculation we confirm that
\begin{eqnarray}
\langle\sigma_{i}^{x}\sigma_{i+1}^{y}\rangle&=&
\langle\sigma_{i}^{y}\sigma_{i+1}^{x}\rangle, \nonumber \\
\langle\sigma_{i}^{z}\sigma_{i+1}^{z}\rangle&=&
\langle \sigma_{i}^{z} \rangle^2              \nonumber \\
&-&\langle\sigma_{i}^{x}\sigma_{i+1}^{x}\rangle\langle\sigma_{i}^{y}\sigma_{i+1}^{y}\rangle
+\langle\sigma_{i}^{x}\sigma_{i+1}^{y}\rangle\langle\sigma_{i}^{y}\sigma_{i+1}^{x}\rangle.
\nonumber
\end{eqnarray}%


\begin{references}


\bibitem{Nus15} Z. Nussinov and J. van den Brink,
                   Rev. Mod. Phys. \textbf{87}, 1 (2015).

\bibitem{Brz07} W. Brzezicki, J. Dziarmaga, and A. M. Ole\'s,
                   Phys. Rev. B \textbf{75}, 134415 (2007);
                W. Brzezicki and A. M. Ole\'s,
                   Acta Phys. Polon. A \textbf{115}, 162 (2009).

\bibitem{Goff95} J. P. Goff, D. A. Tennant, and S. E. Nagler,
                   Phys. Rev. B \textbf{52}, 15992 (1995).

\bibitem{Masuda04} T. Masuda, A. Zheludev, A. Bush, M. Markina, and A. Vasiliev,
                   Phys. Rev. Lett. \textbf{92}, 177201 (2004).

\bibitem{Rusydi08} A. Rusydi, I. Mahns, S. Muller, M. Rubhausen, S. Park,
                   Y. J. Choi, C. L. Zhang, S.-W. Cheong, S. Smadici,
                   P. Abbamonte, M. V. Zimmermann, and G. A. Sawatzky,
                   Appl. Phys. Lett. \textbf{92}, 262506 (2008).

\bibitem{Capogna10} L. Capogna, M. Reehuis, A. Maljuk, R. K. Kremer,
                   B. Ouladdiaf, M. Jansen, and B. Keimer,
                   Phys. Rev. B \textbf{82}, 014407 (2010).

\bibitem{Suzuki71} M. Suzuki,
                   Prog. Theor. Phys. {\bf 46}, 1337 (1971).

\bibitem{Peng09} X. Peng, J. Zhang, J. Du, and D. Suter,
                   Phys. Rev. Lett. \textbf{103}, 140501 (2009).

 \bibitem{Tseng99} C. H. Tseng, S. Somaroo, Y. Sharf, E. Knill,
                   R.~Laflamme, T. F. Havel, and D. G. Cory,
                   Phys. Rev. A \textbf{61}, 012302 (1999).

\bibitem{Zhang06} J. Zhang, X. Peng, and D. Suter,
                   Phys. Rev. A \textbf{73}, 062325 (2006).

\bibitem{Derzhko11} V. Derzhko, O. Derzhko, and J. Richter,
                   Phys. Rev. B \textbf{83}, 174428 (2011).

\bibitem{Eloy12} D. Eloy and J. C. Xavier,
                   Phys. Rev. B  \textbf{ 86}, 064421 (2012).

\bibitem{Titvinidze03} I. Titvinidze and G. I. Japaridze,
                   Eur. Phys. J. B \textbf{32}, 383 (2003).

\bibitem{Topilko12} M. Topilko, T. Krokhmalskii, O. Derzhko, and V. Ohanyan,
                   Eur. Phys. J. B {\bf 85}, 278 (2012).


\bibitem{Men15} O. Menchyshyn, V. Ohanyan, T. Verkholyak, T.~Krokhmalskii,
                   and O. Derzhko,
                   Phys. Rev. B \textbf{92}, 184427 (2015).

\bibitem{Kro08} T. Krokhmalskii, O. Derzhko, J. Stolze, and T.~Verkholyak,
                   Phys. Rev. B \textbf{77}, 174404 (2008).

\bibitem{Lou04} P. Lou, W.-C. Wu, and M.-C. Chang,
                   Phys. Rev. B {\bf 70}, 064405 (2004).

\bibitem{Cheng10b} W. W. Cheng and J.-M. Liu,
                   Phys. Rev. A  \textbf{82}, 012308 (2010).

 \bibitem{Zvyagin09} A. A. Zvyagin,
                   Phys. Rev. B \textbf{80}, 014414 (2009).

\bibitem{Lian11} H. L. Lian and D. P. Tian,
                   Phys. Lett. A \textbf{375}, 3604 (2011).

\bibitem{Cheng10c} W. W. Cheng, C. J. Shan, Y. X. Huang, T. K. Liu, and H. Li,
                   Physica B {\bf 405}, 4821 (2010).

\bibitem{Zhang15} G. Zhang and Z. Song,
                   Phys. Rev. Lett. \textbf{115}, 177204 (2015).

\bibitem{Li11}  Yan-Chao Li and Hai-Qing Lin,
                   Phys. Rev. A \textbf{83}, 052323 (2011).

\bibitem{Lian11b} HanLi Lian, Physica B \textbf{406}, 4278 (2011).

\bibitem{Cheng10a} W. W. Cheng and J.-M. Liu,
                   Phys. Rev. A \textbf{81}, 044304 (2010).

\bibitem{Lou05} Ping Lou,
                   Phys. Rev. B \textbf{72}, 064435 (2005).

\bibitem{Lei15} S. Lei and P. Tong,
                   Physica B \textbf{463}, 1 (2015).

\bibitem{Liu12} Xiaoxian Liu, Ming Zhong, Hui Xu and Peiqing Tong,
                   J. Stat. Mech. P01003 (2012).

\bibitem{Kopp05} A. Kopp and S. Chakravarty,
                   Nat. Phys. {\bf  1}, 53 (2005).

\bibitem{Niu12} Y. Niu, S. B. Chung, C.-H. Hsu, I. Mandal, S. Raghu,
                   and S.~Chakravarty,
                   Phys. Rev. B \textbf{85}, 035110 (2012).

\bibitem{Rajak07} Atanu Rajak and Uma Divakaran,
                   J. Stat. Mech. P04023 (2007).

\bibitem{Dong16} Y. L. Dong, T. Neupert, R. Chitra, and S. Schmidt
                   Phys. Rev. B \textbf{94}, 035441 (2016).

\bibitem{Fradkin78} E. Fradkin and L. Susskind,
                   Phys. Rev. D \textbf{17}, 2637 (1978).

\bibitem{Mochizuki11} M. Mochizuki, N. Furukawa, and N. Nagaosa,
                   Phys. Rev. Lett. \textbf{105}, 037205 (2010);
                   Phys. Rev. B \textbf{84}, 144409 (2011).

\bibitem{You1}  W.-L. You, P. Horsch, and A. M. Ole\'s,
                   Phys. Rev. B \textbf{89}, 104425 (2014).

\bibitem{You2}  W.-L. You, G.-H. Liu, P. Horsch, and A. M. Ole\'s,
                   Phys. Rev. B \textbf{90}, 094413 (2014).

\bibitem{You16} W.-L. You, Y.-C. Qiu, and A. M. Ole\'s,
                   Phys. Rev. B \textbf{93}, 214417 (2016).

\bibitem{Wu17} Q.-C Wu, W.-H Ni, and W.-L. You,
                   J. Phys.: Condens. Matter \textbf{29}, 225804 (2017).

\bibitem{Hirsch79} J. E. Hirsch and G. F. Mazenko,
                   Phys. Rev. B \textbf{19}, 2656  (1979).

\bibitem{Pachos04} J. K. Pachos and M. B. Plenio,
                   Phys. Rev. Lett. {\bf 93}, 056402 (2004).

\bibitem{Qiu16} Y.-C. Qiu, Q.-Q. Wu and W.-L. You,
                   J. Phys.: Condens. Matter \textbf{28}, 496001 (2016).

\bibitem{Robin16} R. Steinigeweg and W. Brenig,
                   Phys. Rev. B \textbf{93}, 214425 (2016).

\bibitem{Zotos97} X. Zotos, F. Naef, and P. Prelov\v{s}ek,
                   Phys. Rev. B \textbf{55}, 11029 (1997).

\bibitem{Antal97} T. Antal, Z. R\'{a}cz, and L. Sasv\'{a}ri,
                    Phys. Rev. Lett. \textbf{78}, 167 (1997).

\bibitem{Salathe15} Y. Salath\'{e}, M. Mondal, M. Oppliger, J. Heinsoo, P. Kurpiers,
                    A. Potocnik, A. Mezzacapo, U. Las Heras, L. Lamata, E. Solano,
                    S. Filipp, and A. Wallraff,
                    Phys. Rev. X \textbf{5}, 021027 (2015).

\bibitem{Mezzacapo14} A. Mezzacapo, L. Lamata, S. Filipp, and E. Solano,
                    Phys. Rev. Lett. \textbf{113}, 050501 (2014).

\bibitem{Bardyn12} C.-E. Bardyn and A. \.{I}mamo\v{g}lu,
                    Phys. Rev. Lett. \textbf{109}, 253606  (2012).

\bibitem{lieb61} E. H. Lieb, T. Schulz, D.Mattis,
                    Ann. Phys. (N.Y.) \textbf{16}, 407 (1961).

\bibitem{Katsura62} S. Katsura,
                    Phys. Rev. \textbf{127}, 1508 (1962).

\bibitem{EBarouch70} E. Barouch and B. M. McCoy,
                    Phys. Rev. A  \textbf{2}, 1075 (1970);
                                  \textbf{3},  786 (1971).

\bibitem{Tong15} Y. Xiong and P. Tong,
                   New J. Phys. \textbf{17}, 013017 (2015);
                X. Wang, T. Liu, and Y. Xiong and P. Tong,
                   Phys. Rev. A \textbf{92}, 012116 (2015).

\bibitem{Jafari17} R. Jafari and H. Johannesson,
                   Phys. Rev. Lett. \textbf{118}, 015701 (2017).

\bibitem{Cin10} L. Cincio, J. Dziarmaga, and A. M. Ole\'s,
                   Phys. Rev. B \textbf{82}, 104416 (2010).

\bibitem{Derzhko98} O. Derzhko,
                   in: {\it Condensed Matter Physics in the Prime of the 21st Century.
                   Phenomena, Materials, Ideas, Methods}, edited by J. Jedrzejewski
                   (World Scientific, Singapore, 2008).

\bibitem{You08} W.-L. You and G.-S. Tian,
                   Phys. Rev. B \textbf{78}, 184406 (2008).

\bibitem{Chiu16} C.-K. Chiu, J. C. Y. Teo, A. P. Schnyder, and S. Ryu,
                   Rev. Mod. Phys. \textbf{88}, 035005 (2016).

\bibitem{Kitaev01} A. Yu. Kitaev,
                   Phys.-Usp. (Suppl.) \textbf{44}, 131 (2001).

\bibitem{Ghosh10} P. Ghosh, J. D. Sau, S. Tewari, and S. Das Sarma,
                   Phys. Rev. B \textbf{82}, 184525 (2010).

\bibitem{Dender97} D. C. Dender, P. R. Hammar, D. H. Reich, C. Broholm,
                   and G. Aeppli,
                   Phys. Rev. Lett. \textbf{79}, 1750 (1997).

\bibitem{Kono15} Y. Kono, T. Sakakibara, C. P. Aoyama, C. Hotta,
                   M. M. Turnbull, C. P. Landee, and Y. Takano,
                   Phys. Rev. Lett. \textbf{114}, 037202 (2015).

\bibitem{Liang15} T. Liang, S. M. Koohpayeh, J. W. Krizan, T. M. McQueen,
                   R. J.  Cava, and N. P. Ong,
                   Nature Commun. \textbf{6}, 7611 (2015).

\bibitem{Sun09a} Ke-Wei Sun and Qing-Hu Chen,
                   Phys. Rev. B {\bf 80}, 174417 (2009).

\bibitem{Motamedifar13} M. Motamedifar, S. Mahdavifar, S. F. Shayesteh, and S. Nemati,
                   Phys. Scr. \textbf{88}, 015003  (2013).

\bibitem{Eri09} E. Eriksson and H. Johannesson,
                   Phys. Rev. B \textbf{79}, 224424 (2009).

\bibitem{Imada1999} M. Imada, F. F. Assaad, H. Tsunetsugu, and Y. Motome,
                   in: \textit{Physics and Chemistry of Transition Metal Oxides},
                   edited by H. Fukuyama and N. Nagaosa
                   (Springer, Berlin, 1999), p. 120-135.

\bibitem{Huijse15} L. Huijse, B. Bauer, and E. Berg,
                   Phys. Rev. Lett. \textbf{114}, 090404 (2015).

\bibitem{Osterloh02} A. Osterloh, L. Amico, G. Falci, and R. Fazio,
                   Nature \textbf{416}, 608 (2002).

\bibitem{Gu04}  Shi-Jian Gu, Shu-Sa Deng, You-Quan Li, and Hai-Qing Lin,
                   Phys. Rev. Lett. \textbf{ 93}, 086402 (2004).

\bibitem{Zanardi07} P. Zanardi, H. T. Quan, X. Wang, and C. P. Sun,
                   Phys. Rev. A {\bf 75}, 032109 (2007).

\bibitem{You07} W.-L. You, Y.-W. Li, and S.-J. Gu,
                   Phys. Rev. E {\bf 76}, 022101 (2007).

\bibitem{Venuti07} L. Campos Venuti and P. Zanardi,
                   Phys. Rev. Lett. {\bf 99}, 095701 (2007).

\bibitem{Gu10}  Shi-Jian Gu,
                   Int. J. Mod. Phys. B {\bf 24}, 4371 (2010).

\bibitem{Miyaji15} M. Miyaji, T. Numasawa, N. Shiba, T. Takayanagi,
                   and K. Watanabe,
                   Phys. Rev. Lett {\bf 115}, 261602 (2015).

\bibitem{Yang07} M.-F. Yang,
                   Phys. Rev. B {\bf 76}, 180403 (2007).

\bibitem{Wang10} B. Wang, M. Feng, and Z.-Q. Chen,
                   Phys. Rev. A {\bf 81}, 064301 (2010).

\bibitem{Sun15} G. Sun, A. K. Kolezhuk, and T. Vekua,
                   Phys. Rev. B {\bf 91}, 014418 (2015).

\bibitem{You15} W.-L. You and L. He,
                   J. Phys.: Condens. Matter  \textbf{27}, 205601 (2015).

\bibitem{Liu11} B.-Q. Liu, B. Shao, J.-G. Li, J. Zou, and L.-Ao Wu,
                   Phys. Rev. A \textbf{83}, 052112 (2011).

\bibitem{Chen16}Jin-Jun Chen, Jian Cui, Yu-Ran Zhang, and Heng Fan,
                   Phys. Rev. A \textbf{94}, 022112 (2016).

\bibitem{Dil08} R. Dillenschneider,
                   Phys. Rev. B \textbf{78}, 224413 (2008).

\bibitem{Sarandy09} M. S. Sarandy,
                   Phys. Rev. A \textbf{80}, 022108 (2009).

\bibitem{Werlang10} T. Werlang, C. Trippe, G. A. P. Ribeiro, and G. Rigolin,
                   Phys. Rev. Lett. \textbf{105}, 095702 (2010).

\bibitem{Ollivier01} H. Ollivier and W. H. Zurek,
                   Phys. Rev. Lett. \textbf{88}, 017901 (2001).

\bibitem{Wootters98} W. K. Wootters,
                   Phys. Rev. Lett. \textbf{80}, 2245 (1998).

\end{references}
\end{document}